\documentclass[useAMS,usenatbib]{mnras}
\usepackage{hyperref}
\usepackage{amsmath}
\usepackage{amssymb}
\usepackage{graphicx}

\def\d{{\rm d}}
\def\e{{\rm e}}
\def\i{{\rm i}}
\def\Re{\mathop{\hbox{Re}}}
\def\mass{M}
\def\pol{P}
\def\Im{\mathop{\hbox{Im}}}

\def\vsc{v_{\rm s}}
\def\i{{\rm i}}
\def\fzero{F}
\def\stim{^{\rm e}}
\def\resp{^{f}}
\def\pert{_{\rm p}}
\def\PV{{\cal P}}

{\newif\ifnotend
\notendtrue
\def\veclist{ABCDEFGHIJKLMNOPQRSTUVWXYZabcdfghijklmnopqrstuvwxyz.}
\def\top#1#2.{#1}
\def\tail#1#2.{#2.}
\loop\expandafter\xdef\csname v\expandafter\top\veclist\endcsname%
{{\noexpand\bf\expandafter\top\veclist}}
\edef\veclist{\expandafter\tail\veclist}
\if\veclist.\notendfalse\fi\ifnotend\repeat}

\def\vzero{{\bf 0}}
\let\boldgrk=\gkvecten
\let\boldgrksc=\gkvecseven

\def\gkthing#1{{\mathchoice%
    {\hbox{{\boldgrk\char#1}}}
    {\hbox{{\boldgrk\char#1}}}
    {\hbox{{\boldgrksc\char#1}}}
    {\hbox{{\boldgrksc\char#1}}}}}

\def\vtheta{\gkthing{18}}

\def\vOmega{\gkthing{10}}

\begin{document}
\title[Periodic cube]{Stellar dynamics in the periodic cube}
\author[SJ Magorrian]{John Magorrian\\
Rudolf Peierls Centre for Theoretical Physics, Clarendon Laboratory,
Parks Road, Oxford OX1 3PU}

\date{Accepted 2021 August 24. Received 2021 August 9; in original form 2021 June 18}

\maketitle

\label{firstpage}

\begin{abstract}
  We use the problem of dynamical friction within the periodic cube to
  illustrate the application of perturbation theory in stellar
  dynamics, testing its predictions against measurements from $N$-body
  simulation.
  Our development is based on the explicitly time-dependent Volterra
  integral equation for the cube's linear response, which
  avoids the subtleties encountered in analyses based on complex
  frequency.
  We obtain an expression for the self-consistent response of the cube
  to steady stirring by an external perturber.
  From this we show how to obtain the familiar Chandrasekhar dynamical
  friction formula and construct an elementary derivation of the
  Lenard--Balescu equation for the secular quasilinear evolution of an
  isolated cube composed of $N$ equal-mass stars.
  We present an alternative expression for the (real-frequency) van
  Kampen modes of the cube and show explicitly how to decompose any
  linear perturbation of the cube into a superposition of such modes.
\end{abstract}


\section{Introduction}

The periodic cube is an extremely simple model of a self-gravitating
stellar system.  Stars are confined to the cube because space is
assumed to be periodic: for any function $g$ of the spatial
coordinates $(x,y,z)$ in a cube of side~$L$ we have that
$g(x+L,y,z)=g(x,y+L,z)=g(x,y,z+L)=g(x,y,z)$.  The simplest, most
natural equilibrium models have uniform mass density~$\rho_0$ and
vanishing gravitational potential, $\Phi(\vx)=0$, so that stars'
orbits are straight lines.

The periodic cube is sometimes encountered en route to elementary
derivations of the behaviour of infinite homogeneous stellar systems,
but was first explicitly identified by
\cite{BarnesDynamicalinstabilitiesspherical1986}
and then studied in more detail by
\cite{WeinbergNonlocalcollectiverelaxation1993} and \cite{AubertMesureimplicationsdynamiques2005}.  It has received
little attention beyond that, no doubt because it is such an
unrealistic model galaxy.  Nevertheless, its simplicity makes it a
useful backdrop to illustrate the application of perturbation theory
to stellar dynamics.

One of the most obvious applications of perturbation theory is in
understanding the process of dynamical friction: a massive body
passing through or nearby a stellar system induces a density wake; in
most cases the gravitational force from this wake slows the body.
This is a well-studied, classic problem
\citep[e.g.,][]{ChandrasekharDynamicalFrictionGeneral1943,
  WeinbergOrbitaldecaysatellite1986,Maozfluctuationdissipationapproach1993,ChavanisKinetictheoryspatially2013b},
but most investigations have ignored the wake's self-gravity when
computing the response: stars are treated as massless test particles
moving in the combined potential of the perturbing massive body and
the unperturbed stellar system, which results in an incomplete account
of the response.  In the case of a satellite orbiting a spherical
galaxy, \cite{WhiteSimulationssinkingsatellites1983} and
\cite{WeinbergSelfgravitatingresponsespherical1989} demonstrate that
neglecting self gravity leads to satellite sinking rates that are in
error by a factor of 2 or more.

In this paper we use linear perturbation theory to study dynamical
friction in the periodic cube in some detail, focusing on the time
dependence of the response and on the structure of the wakes, both
with and without self gravity.
In Section~2 we introduce some simple equilibrium cube models and show
how to calculate the linear response to an externally applied
perturbation.  The fundamentals of this calculation are the same as
for any other stellar system, but a special property of the cube is
that its evolution equations are particularly easy to solve when
Fourier transformed in velocity, a feature we exploit unrelentingly
throughout the paper.
We use this formalism in Section~3 to calculate the dynamical friction
force felt by a massive body moving through the cube and compare the
results to measurements from $N$-body simulation.
In Section~4 we use these results to show how to decompose a
perturbation of an isolated cube into a superposition of indepdendent
undamped oscillations -- its \cite{vanKampentheorystationarywaves1955}
modes.
Then, stepping slightly beyond the remit of purely linear perturbation
theory, in Section~5 we use the results from Section~3 to derive
the Lenard--Balescu equation
\citep{BalescuIrreversibleProcessesIonized1960,LenardBogoliubovkineticequation1960}
for the quasilinear secular evolution of $N$-body realisations of
isolated cubes and test whether its predictions agree with
measurements from simulations.
Section~6 sums up.  An appendix gives details of our $N$-body
simulations.

\begin{table*}
	\centering
	\label{tab:fzeros}
	\begin{tabular}{llll} 
		\hline
          & $\fzero(\vv)$ & $\mathop{\hbox{Var}}\fzero(v_i)$ &$\tilde\fzero(\vk)$ \\
          \hline
          Maxwellian
          &
          $\frac1{(2\pi\vsc^2)^{3/2}L^3}\exp\left[-\frac{\vv^2}{2\vsc^2}\right]$
          &
            $\vsc^2$
            &
            $(2\pi)^{-3}\exp\left[-\frac12\ell^2\vsc^2\vk^2\right]$
          \\
          Long-tailed (Lorentzian)
          &
          $\frac\vsc{\pi^2L^3(\vv^2+\vsc^2)^2}$
                          &
                            $\infty$
                          &
                           $(2\pi)^{-3}\exp[-(\ell\vsc)k]$\\
          Isotropic top hat
          &
            $\frac1{\frac43\pi\vsc^3L^3}\Theta\left(1-\frac{|\vv|}\vsc\right)
            $
            & $\frac15\vsc^2$
          &
            $\frac3{(2\pi k\ell\vsc)^3}\left[\sin k\ell\vsc-k\ell\vsc\cos
            k\ell\vsc\right]
            =-\frac3{(2\pi)^3k\ell\vsc}\frac{\d}{\d(k\ell\vsc)}\frac{\sin k\ell\vsc}{k\ell\vsc}$
          \\
          Cubic top hat
          &
            $\prod_{i=1}^3\frac1{2L\vsc}\Theta\left(1-\frac{|v_i|}\vsc\right)$
            & $\frac13\vsc^2$
                          &
                            $\prod_{i=1}^3
                            \frac1{2\pi}\frac{\sin(k_i\ell\vsc)}{k_i\ell\vsc}$
          \\
          \hline
	\end{tabular}
	\caption{The equilibrium DFs $F(\vv)$ used in this paper
          together with their (one-dimensional) second moments and
          Fourier transforms $\tilde\fzero(\vk)$
          (equation~\ref{eq:FT}).  Each DF has a
          single parameter, $\vsc$, that controls the characteristic
          width of the velocity distribution.
          The Heaviside function $\Theta(x)$ used in the definition of
          the top-hat DFs returns 1 when $x\ge0$
          and 0 otherwise.  } 
\end{table*}

\begin{table*}
	\centering
	\label{tab:transforms}
        \begin{equation*}
          \vtheta \equiv \vx/\ell,\quad \vJ\equiv\ell \vv,\quad \tau\equiv
          t/\ell^2,\quad \vOmega\equiv\vJ/\ell^2.
        \end{equation*}
        \begin{equation*}
          \hbox{Angle}\qquad
          g_\vn(\vJ)\equiv\frac1{(2\pi)^3}\int \e^{-\i\vn\cdot\vtheta}g(\vtheta,\vJ)\d^3\vtheta;
        \qquad
        g(\vtheta,\vJ)=\sum_{\vn}g_\vn(\vJ)\e^{\i\vn\cdot\vtheta}.
      \end{equation*}
        \begin{equation*}
          \hbox{Action}\qquad
        \tilde g_\vn(\vk)\equiv\int\e^{-\i\vk\cdot\vJ}g_\vn(\vJ)\,\d^3\vJ;\qquad
        g_\vn(\vJ)=\frac1{(2\pi)^3}\int \e^{\i\vk\cdot\vJ}\tilde g_{\vn}(\vJ)\,\d^3\vk.
        \end{equation*}
	\caption{Summary of the transformations used in this paper.
          The length scale $\ell=L/2\pi$ (equation~\ref{eq:ell}), where $L$ is the length of
          the cube.}
 \end{table*}

\section{The periodic cube and its response to perturbations}

Our unperturbed cube is perfectly collisionless.  Stars' orbits are
straight lines.  The velocity~$\vv$ is then an integral of
motion, and, by Jeans' theorem, the equilibrium probability density
distribution of stars (DF), $F(\vx,\vv)$ must be a function $F=F(\vv)$
of~$\vv$ only.  There are no unbound stars in the cube, which means
that any $F(\vv)$ is an acceptable equilibrium DF, albeit not
necessarily a stable equilibrium.

In this paper we consider the four DFs listed in Table~\ref{tab:fzeros}.
Each has a single parameter~$\vsc$ that controls the width of the
velocity distribution.  The Maxwellian DF is familiar from the kinetic
theory of gases.  Stellar systems are in general not completely
relaxed, however, so we also consider additional DFs that ``bracket''
what one might reasonably expect to find in a galaxy.  The long-tailed
DF \citep[introduced by][]{Summersmodifiedplasmadispersion1991} is so
extended that its second moment does not exist.  The marginal,
one-dimensional velocities in this model are Lorentzian,
$f(v_i)\propto 1/(v_i^2+\vsc^2)$.  In constrast, the isotropic top-hat
DF is as concentrated as possible for its velocity dispersion without
having $\partial F/\partial|\vv|<0$.  The cubic top-hat DF is similar,
but has an anisotropic velocity distribution aligned with the
edges of the cube.

\subsection{Perturbation expansion}

Now consider what happens when an external potential perturbation
$\epsilon\Phi\stim(\vx,t)$ is applied to the cube.
In response the DF changes from $\fzero$ to $\fzero+\epsilon
f(\vx,\vv,t)$, which has corresponding potential
$\epsilon\Phi\resp$.
The evolution of the DF is described by the CBE,
\begin{equation}
  \begin{split}
  \label{eq:fullCBExv}
  &\epsilon^0\left[
    \frac{\partial F}{\partial t}
    +\vv\cdot\frac{\partial  F}{\partial\vx}
  \right]
  +\epsilon^1\left[
    \frac{\partial f}{\partial t}
    +\vv\cdot\frac{\partial f}{\partial\vx}
    -\frac{\partial\Phi}{\partial\vx}\cdot\frac{\partial F}{\partial\vv}
  \right]\\
  &\quad
  +\epsilon^2\left[
    -\frac{\partial
      f}{\partial\vv}\cdot\frac{\partial\Phi}{\partial\vx}
    \right]=0,
  \end{split}
\end{equation}
in which the potential
\begin{equation}
  \Phi=\Phi\stim+\Phi\resp
  \label{eq:Phisplit}
\end{equation}
is the sum of stimulus and response potentials, with the latter
depending on the response density $\rho\resp(\vx)=M\int f(\vx,\vv)\,\d^3\vv$.

We follow the usual procedure of rewriting the
CBE~\eqref{eq:fullCBExv} in terms of
angle--action variables $(\vtheta,\vJ)$.
For the periodic cube the
angle--action coordinates are obtained directly from $(\vx,\vv)$ by
the trivial rescaling
\begin{equation}
  \label{eq:aa}
  \vtheta = \vx/\ell,\quad \vJ=\ell \vv,
\end{equation}
where
\begin{equation}
  \label{eq:ell}
  \ell\equiv \frac L{2\pi}.
\end{equation}
As the coordinates $\vtheta$ are $2\pi$ periodic, any function $g(\vtheta,\vJ)$
can be expressed as the Fourier series
\begin{equation}
  g(\vtheta,\vJ)=\sum_\vn g_\vn(\vJ)\e^{\i\vn\cdot\vtheta},
\label{eq:Fseries}
\end{equation}
where
the sum is over triplets of integers $\vn=(n_1,n_2,n_3)$, with coefficients
\begin{equation}
  g_\vn(\vJ)=\frac1{(2\pi)^3}\int
  \e^{-\i\vn\cdot\vtheta}\,g(\vtheta,\vJ)\,\d^3\vtheta.
\label{eq:Fcoeff}
\end{equation}
From~\eqref{eq:fullCBExv}, the Fourier spatial modes $f_\vn(\vJ,t)$ of the DF response obey
\begin{equation}
  \begin{split}
  \label{eq:fullCBE}
    & \epsilon\left[\frac{\partial f_\vn}{\partial t}+\i\vn\cdot\vOmega f_\vn
  -\i\vn\cdot\frac{\partial \fzero}{\partial \vJ}\Phi_\vn\right]\\
&\quad+\epsilon^2\left[
  \sum_\vm-\i\vm\cdot\frac{\partial
      f_{\vn-\vm}}{\partial\vJ}\Phi_\vm
\right]  =0,
  \end{split}
\end{equation}
where $\vOmega\equiv\vJ/\ell^2$ is the vector of frequencies of an orbit
with action~$\vJ$.
Later it will prove convenient to introduce a rescaled time variable
\begin{equation}
  \label{eq:tau}
  \tau\equiv t/\ell^2,
\end{equation}
so that $\vOmega t=\vJ\tau$.

Similarly, the potential can be expanded as
$\Phi(\vtheta,t)=\sum_\vn\Phi_\vn(t)\e^{\i\vn\cdot\vtheta}$.
We assume that interaction among stars are invariant under
translations in~$\vx$.  Therefore the Fourier coefficients of the potential
must be given by
\begin{equation}
  \Phi_\vn(t)=V_\vn\rho_\vn
  \label{eq:Vnprop}
\end{equation}
for some set of constants~$V_\vn$, where $\rho_\vn$ are the Fourier
coefficients~\eqref{eq:Fcoeff} of the mass density distribution $\rho(\vx)$.
Conservation of mass requires that
$V_\vn=0$ for $\vn=(0,0,0)$, but the $V_\vn$ are otherwise arbitrary.
In this paper we are interested in systems that satisfy the familiar
Poisson equation
\begin{equation}
  \nabla^2\Phi=4\pi G\rho
  \label{eq:poissonn}
\end{equation}
and therefore take
\begin{equation}
  V_\vn=
  \begin{cases}
  -\frac{4\pi G\ell^2}{\vn^2}, & n_{\rm min}\le|\vn|\le n_{\rm max},\cr
    0, &\hbox{otherwise}.
  \end{cases}
\label{eq:Vn}
\end{equation}
The restrictions $n_{\rm min}\ge1$ and $n_{\rm max}$ on the range of values of
$|\vn|$ for which the potential is ``active'' allows us
to isolate different spatial frequencies in the response

The contribution of the DF response $f(\vx,\vv)$ to the density
$\rho(\vx)$ is simply $\rho\resp(\vx)=M\int f(\vx,\vv)\,\d^3\vv$.  Then,
from~\eqref{eq:Vnprop}, we have that
\begin{equation}
 \Phi\resp_\vn=\frac{MV_\vn}{\ell^3}\int
f_\vn(\vJ)\d^3\vJ. 
\label{eq:fn2Phin}
\end{equation}
That is, 
each $\Phi\resp_\vn$ depends only on the
corresponding~$f_\vn$: this is one of the two immediate major
simplifications introduced for the special case of the periodic cube,
the other one being simply that angle--action variables are trivial to
construct.  
The independence of $\Phi\resp$ on $\vJ$ has led to a minor
simplification of the nonlinear $O(\epsilon^2)$ term in the
CBE~\eqref{eq:fullCBE}.  Apart from that, this equation looks identical to
the CBE for more realistic galaxy models.
But it turns out that having
$\Omega(\vJ)\propto\vJ$ in the linear $O(\epsilon)$ term allows a
powerful further simplification, which we now exploit.

\subsection{Evolution of perturbations}

The usual way of solving~\eqref{eq:fullCBE} to find $f_\vn(\vJ,t)$ and
$\Phi\resp_\vn(t)$ is by
considering only the linearized equation,
\begin{equation}
   \label{eq:linCBE}
   \frac{\partial f_\vn}{\partial t}+\i\vn\cdot\vOmega f_\vn
   -\i\vn\cdot\frac{\partial \fzero}{\partial \vJ}\Phi_\vn=0,
\end{equation}
and then Laplace transforming in time.  This yields a simple algebraic
expression for (the Laplace transform of) the response $f$.  This
expression, however, is valid only in the upper half complex plane of
the transformed time variable, which in practice makes it difficult to
invert to find the response.

As an alternative, let us Fourier transform~\eqref{eq:fullCBE}
in~$\vJ$, defining
\begin{equation}
  \tilde
f_\vn(\vk)\equiv\int\e^{-\i\vk\cdot\vJ}f_\vn(\vJ)\,\d^3\vJ,
\label{eq:FT}
\end{equation}
and similarly for $\tilde\fzero(\vJ)$.  
Applying this Fourier transform to the CBE~\eqref{eq:fullCBE} gives
\begin{equation}
  \begin{split}
    &\epsilon\left[\frac{\partial\tilde f_\vn}{\partial t}
  -\frac1{\ell^2}\vn\cdot\frac{\partial \tilde f_\vn}{\partial\vk}
  +\vn\cdot\vk\,\tilde\fzero\Phi_{\vn}\right]\\
&\quad+\epsilon^2\left[\sum_\vm\vm\cdot\vk\,\tilde
  f_{\vn-\vm}\Phi_\vm\right]
  =0,
  \end{split}
\label{eq:kCBE}
\end{equation}
which, like~\eqref{eq:fullCBE}, is a first-order quasilinear PDE for $\tilde f_\vn(\vk,t)$.  As
such, it is easily solved using the method of characteristics
\citep[e.g.,][]{ArnoldOrdinaryDifferentialEquations1992}.  The
characteristic curves for~\eqref{eq:kCBE} are the straight lines
\begin{equation}
  \label{eq:characteristic}
  \vk(\tau)=\vk(0)-\tau\vn,
\end{equation}
parametrized by the rescaled time variable~$\tau$ defined in~\eqref{eq:tau}.
Along any such characteristic, the value of $f_\vn$ is
governed by the ordinary differential equation
\def\func#1{\tilde f_\vn(\vk(#1),#1)}
\def\func#1{\tilde f_\vn(#1)}
\begin{equation}
  \begin{split}
    &
  \epsilon\left[\frac1{\ell^2}\frac{\d\func{\tau}}{\d\tau}
  +\vn\cdot\vk(\tau)
  \tilde
  \fzero(\tau)\Phi_\vn(\tau)\right]\\
  &+\epsilon^2\left[\sum_\vm\vm\cdot\vk(\tau)\tilde
  f_{\vn-\vm}(\tau)\Phi_\vm(\tau)\right]
  =0,
  \end{split}
\label{eq:charode}
\end{equation}
where we have adopted the shorthand notation $\tilde f_\vn(\tau)=\tilde
f_\vn(\vk(\tau),\tau)$ and $\tilde\fzero(\tau)=\tilde
\fzero(\vk(\tau))$.

It is worth pausing to relate these characteristics to those of the
untransformed CBE~\eqref{eq:fullCBExv}.  The latter are simply the
straight-line orbits $\vtheta(\tau)=\vJ\tau+\hbox{const}$ that stars
follow in the unperturbed cube.  To understand~\eqref{eq:charode}, consider the DF
$f(\vtheta,\vJ,0)=\delta(\vJ-\vJ_0)\delta(\vtheta-\vtheta_0)$ of a
cube that contains a single star that sets out from
$(\vtheta_0,\vJ_0)$.  In our Fourier-transformed picture, the Fourier modes of
this initial DF are
$\tilde f_\vn(\vk,0)=\exp[-\i(\vn\cdot\vtheta_0+\vk\cdot\vJ_0)]$.  In
the absence of any perturbations we have from~\eqref{eq:charode} that
$\d\tilde f_\vn/\d\tau=0$ along~\eqref{eq:characteristic}.  So, at
later times the spatial modes of the DF are given by
\begin{equation}
  \label{eq:DF1undressed}
  \tilde f_\vn(\vk,\tau)=\exp[-\i\vk\cdot\vJ_0]\exp[-\i(\vn\cdot(\vtheta_0+\vJ\tau
))],  
\end{equation}
which represent the original DF translated in
angle by $\vJ \tau$.

In introducing the Fourier expansion~\eqref{eq:Fseries} and
transform~\eqref{eq:FT} we have exchanged a single first-order
quasilinear PDE~\eqref{eq:fullCBExv} for another, harder-to-interpret
set of such equations~\eqref{eq:kCBE}.  The benefit of this exchange,
however, comes from the transformed version of Poisson's
equation~\eqref{eq:poissonn}.
Setting $\vk=0$ in~\eqref{eq:FT} and substituting into~\eqref{eq:fn2Phin},
we have that 
\begin{equation}
  \Phi\resp_\vn(\tau)=\frac{MV_\vn}{\ell^3}\tilde f_\vn(\vk=0,\tau).
  \label{eq:poissonk}
\end{equation}

\subsection{Linear response}
\label{sec:linear}

The linearized response to an imposed $\Phi\stim(\vtheta,t)$ is
straightforward to obtain:
integrate~\eqref{eq:charode} along the characteristic
$\vk(\tau)=(\tau-\tau')\vn$ from $\tau'=-\infty$ to $\tau'=\tau$ and then use
equation~\eqref{eq:poissonk} to express the resulting $\tilde f_\vn(\vk=0,\tau)$ in
terms of $\Phi\resp_\vn(\tau)$. The result is that
\begin{equation}
  \begin{split}
    \Phi\resp_\vn(\tau)&=-\frac{\vn^2MV_\vn}\ell\int_{-\infty}^\tau
  (\tau-\tau')
  \Phi_{\vn}(\tau')
  \tilde\fzero\left(\vn(\tau-\tau')\right)\,\d\tau',\\
  \end{split}
\label{eq:Phiresp}
\end{equation}
in which the stimulus $\Phi\stim_\vn(\tau')$ enters through the
$\Phi_\vn(\tau')=\Phi\stim_\vn(\tau')+\Phi\resp_\vn(\tau')$ factor in the integrand.
This is a linear Volterra equation of the second kind, which is easy
to solve numerically.

If we instead integrate~\eqref{eq:charode} along the characteristic
$\vk(\tau')=\vk+(\tau-\tau')\vn$ then we obtain the more general result that
\begin{equation}
  \begin{split}
    &\tilde f_\vn(\vk,\tau)=-\ell^2\times\\
    &\int_{-\infty}^\tau\d\tau'\,\left[\vn\cdot\vk+\vn^2(\tau-\tau')\right]
    \Phi_\vn(\tau')
    \tilde\fzero(\vk+\vn(\tau-\tau')).\\
  \end{split}
  \label{eq:DFresplin}
\end{equation}
Changing variables from $\tau$ back to $t=\tau/\ell^2$ and taking the
inverse Fourier transform of this gives
\begin{equation}
  \begin{split}
    f_\vn(\vJ,t)&=\i\vn\cdot\frac{\partial\fzero}{\partial\vJ}
    \int_{-\infty}^t\d t'\,\e^{-\i\vn\cdot\vOmega(t-t')}\Phi_\vn(t'),
  \end{split}
  \label{eq:DFresponse}
\end{equation}
the result we would have obtained had we simply solved directly the
CBE~\eqref{eq:fullCBE} to linear order.

\subsection{Linear stability}
\label{sec:linstab}

Of course, calculating the response of the cube to an imposed
perturbation is usually of interest only if the cube is stable in the
absence of perturbations.
Suppose that, in the absence of any external perturbation (that is,
$\Phi\stim=0$), the cube supports a self-sustaining perturbation of
the form $\Phi\resp_\vn=\e^{-\i\omega\tau}$ for some (rescaled-by-$\ell^2$) complex
frequency~$\omega$, which we may split into real and imaginary parts:
$\omega=\omega_{\rm R}+\i\gamma$.  If $\gamma$, the imaginary part of
$\omega$, is positive then the cube is unstable: even the slightest
perturbation can excite a wave that grows as $\e^{\gamma\tau}$.

To find the spectrum of values of $\omega$ that the integral
equation~\eqref{eq:Phiresp} admits let us substitute
$\Phi\resp_\vn=\e^{-\i\omega\tau}$ and $\Phi\stim_\vn=0$.
Introducing the function
\begin{equation}
  \label{eq:In}
  \pol_\vn(\omega)\equiv-\frac{\vn^2MV_\vn}\ell\int_0^\infty \tau \e^{\i\omega\tau}\tilde\fzero(\vn\tau)\,\d\tau,
\end{equation}
equation~\eqref{eq:Phiresp} can be written as
\begin{equation}
  \epsilon_\vn(\omega)\equiv 1-\pol_\vn(\omega)=0,
\label{eq:dispersionreln}
\end{equation}
which is the dispersion relation that must be satisfied by~$\omega$.
Following the well-known analogy with electrostatic media
\citep[e.g.,][]{BinneyGalacticDynamicsSecond2008}, the function
$\pol_\vn(\omega)$ is the polarizability of spatial mode $\vn$ to an imposed
wave of frequency~$\omega$.
The dispersion
relation~\eqref{eq:dispersionreln} is satisfied when the permittivity
or dielectric function
of the cube, $\epsilon=1-\pol$, is equal to zero.
The easiest way to check for stability is by
plotting contours of $|1-\pol_\vn(\omega)|$ in the
$(\omega_{\rm R},\gamma)$ plane and using those to identify roots of
the dispersion relation $1-\pol_\vn(\omega)=0$.  If any of the roots lie
in the upper half plane ($\gamma>0$) then the cube is unstable.

Another way of writing~\eqref{eq:In} is by recognizing that the
$\tau\tilde\fzero(\vn\tau)$ factor in the integrand is the Fourier
transform of the derivative of $\fzero(\vJ)$, namely
\begin{equation}
  \pol_\vn(\omega)=-\frac{\vn^2MV_\vn}\ell
  \int_0^\infty\d\tau\,\e^{\i\omega\tau}
  \frac1{\i\vn^2}
  \int\d^3\vJ\,\e^{-\i\vJ\cdot\vn\tau}
  \vn\cdot\frac{\partial\fzero}{\partial\vJ}.
\end{equation}
If $\Im\omega>0$ we may interchange the order of the integrals in this
expression.  Carrying out the $\tau$
integral gives
\begin{equation}
  \pol_\vn(\omega)\equiv-\frac{MV_\vn}\ell
  \int\frac{\d^3\vJ}{\omega-\vn\cdot\vJ}
  \vn\cdot\frac{\partial\fzero}{\partial\vJ},
\label{eq:stdP}
\end{equation}
which is eq.~(5.56) of \cite{BinneyGalacticDynamicsSecond2008}.  For
$\Im\omega\le0$ we need to analytically continue the integrand
of~\eqref{eq:stdP} to complex values of $\vn\cdot\vJ$.  In contrast,
the expression~\eqref{eq:In} can typically be computed directly.

For a Maxwellian DF (Table~\ref{tab:fzeros})
equation~\eqref{eq:In} gives the polarizability as
\footnote{Compare eq.~(5.64) of \cite{BinneyGalacticDynamicsSecond2008}.}
\begin{equation}
  \begin{split}
    \pol_\vn(\omega)
    &=-\frac{MV_\vn}{(2\pi\ell)^3\vsc^2}
  \Bigg[
  1+\i\omega'
  {\sqrt{\frac\pi2}\left(1+\mathop{\hbox{erf}}\left(\frac{\i\omega'}{\sqrt2}\right)\right)
  \e^{-\frac12\omega'^2}}
    \Bigg]
    ,
\end{split}
\label{eq:InMaxwell}
\end{equation}
where the dimensionless frequency
$\omega'\equiv\omega/n\ell\vsc$.
Taking $V_\vn$ from equation~\eqref{eq:Vn} and examining contours of
$|1-\pol_\vn(\omega)|$ is it easy to see that this model is stable
provided
\begin{equation}
  M<M_{\rm J}\equiv\frac{(2\pi)^3\vsc^2\ell n_{\rm min}^2}{4\pi G}.
\label{eq:MaxwellMJ}
\end{equation}
To place this result in context, consider the {\it infinite}
homogeneous system \citep[e.g.,][]{BinneyGalacticDynamicsSecond2008,ChavanisDynamicalstabilityinfinite2012}
with the same Maxwellian DF in velocity.  The infinite system is
unstable to perturbations having wavelengths
$\lambda>\lambda_{\rm J}$, where the Jeans length
$\lambda^2_{\rm J}\equiv \pi\vsc^2/G\rho_0$.  In the periodic cube
the density $\rho_0=M/(2\pi\ell)^3$ and the longest wavelength
oscillation that the cube admits is
$\lambda_{\rm max}=2\pi\ell/n_{\rm min}$.  The
$\lambda>\lambda_{\rm J}$ condition for Jeans stability is our
equation~\eqref{eq:MaxwellMJ}.

For the long-tailed distribution the polarizability is
\footnote{See eq.~(5.204) of \cite{BinneyGalacticDynamicsSecond2008},
  eq.~(94) of \cite{ChavanisLinearresponsetheory2013}.}
\begin{equation}
  \pol_\vn(\omega)=-\frac{MV_\vn}{(2\pi\ell)^3\vsc^2}
  \frac1{\left(1-\i\omega'\right)^2},
\end{equation}
which leads to the same condition~\eqref{eq:MaxwellMJ} for Jeans
stability as in the Maxwellian DF.
For the isotropic top-hat DF the polarizability is
\begin{equation}
  \begin{split}
    \pol_\vn(\omega)
    =-\frac{3MV_\vn}{(2\pi\ell)^3\vsc^2}
    \left[1+\frac12\omega'
      \log\left(\frac{\omega'-1}{\omega'+1}\right)
      \right],
  \end{split}
\end{equation}
while for the cubic top hat, assuming $\Im\omega>0$, it
is\footnote{See eq.~(89) of \cite{ChavanisLinearresponsetheory2013}}
\begin{equation}
  \begin{split}
    \pol_\vn(\omega)
    =-\frac{MV_\vn}{(2\pi\ell)^3\vsc^2}
    \frac1{\omega'^2-1}.
  \end{split}
\end{equation}
By inspection of the latter it is clear that no matter how small $M$
is there will be some $\omega=\omega_{\rm R}+\i0^+$ for which
$1-\pol_\vn(\omega)=0$: that is, the cubic top hat DF always supports
neutrally stable oscillations.  In contrast all linear oscillations of
the isotropic top hat model are damped as long as
\begin{equation}
  M<M_{\rm J}\equiv\frac13\frac{(2\pi)^3\vsc^2\ell n_{\rm min}^2}{4\pi G},
\label{eq:IsoTopHatMJ}
\end{equation}
a slightly different Jeans criterion to that for the Maxwellian and
long-tailed models~\eqref{eq:MaxwellMJ}.

\begin{figure*}
  \centering
  \includegraphics[width=0.32\hsize]{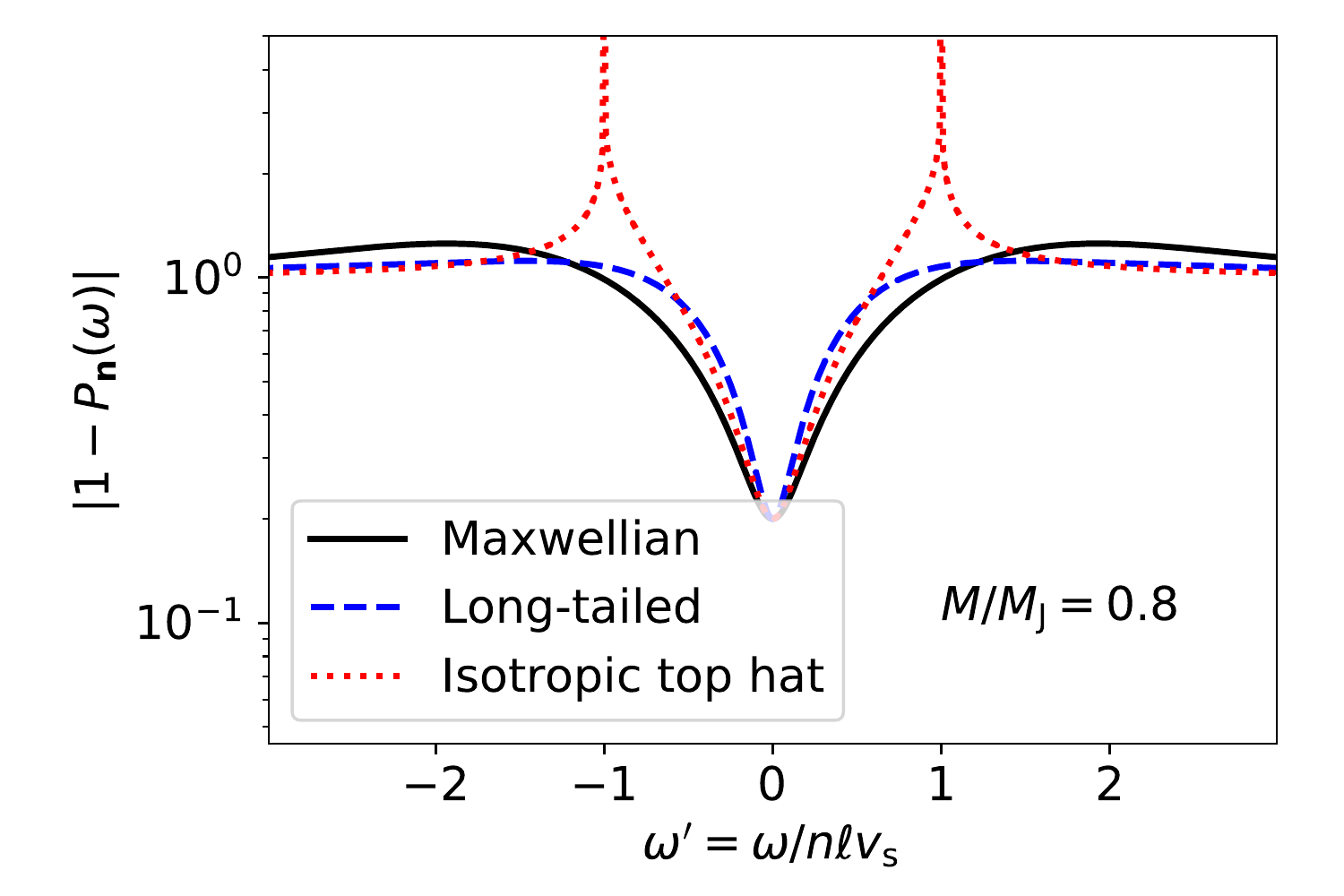}
  \includegraphics[width=0.32\hsize]{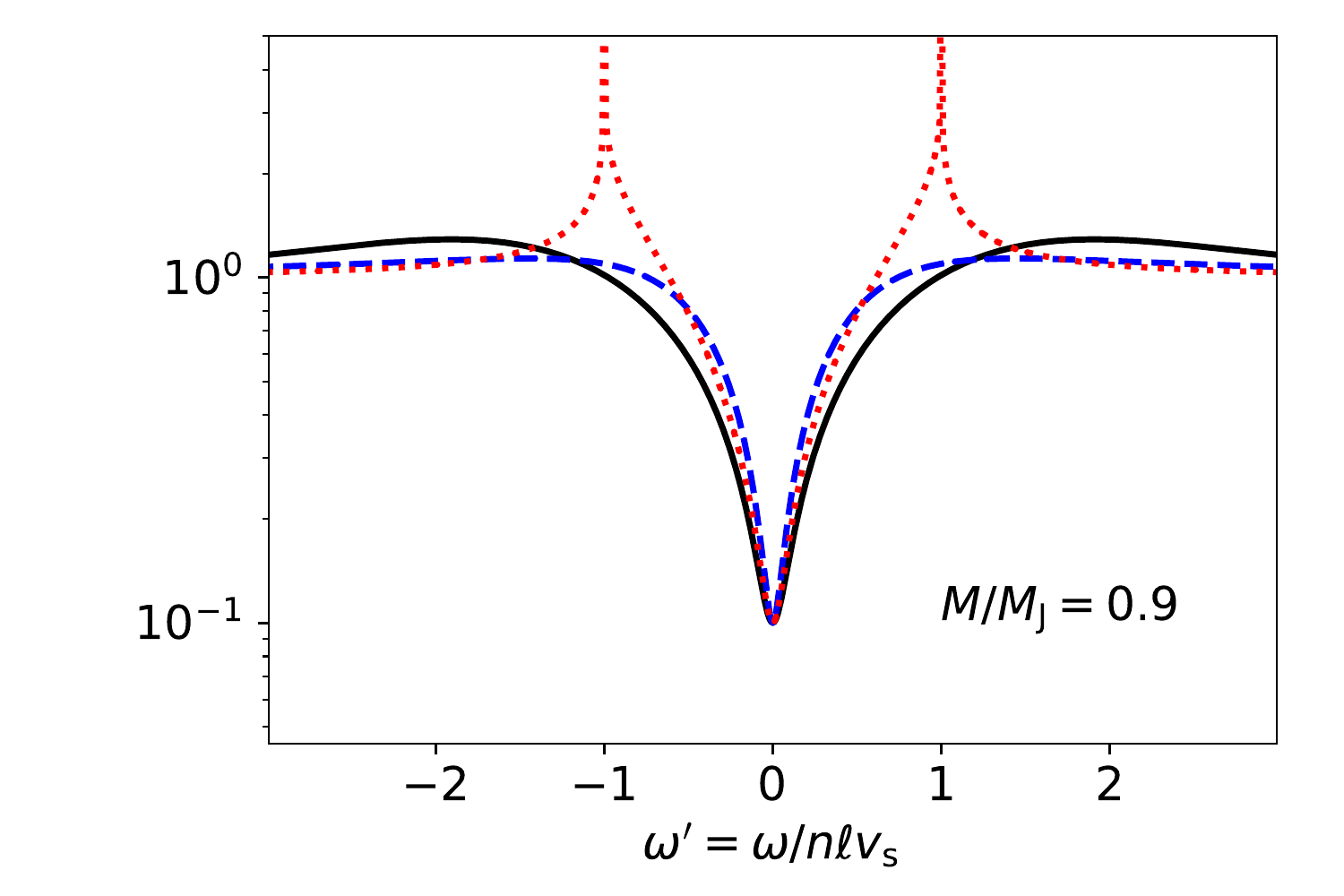}
  \includegraphics[width=0.32\hsize]{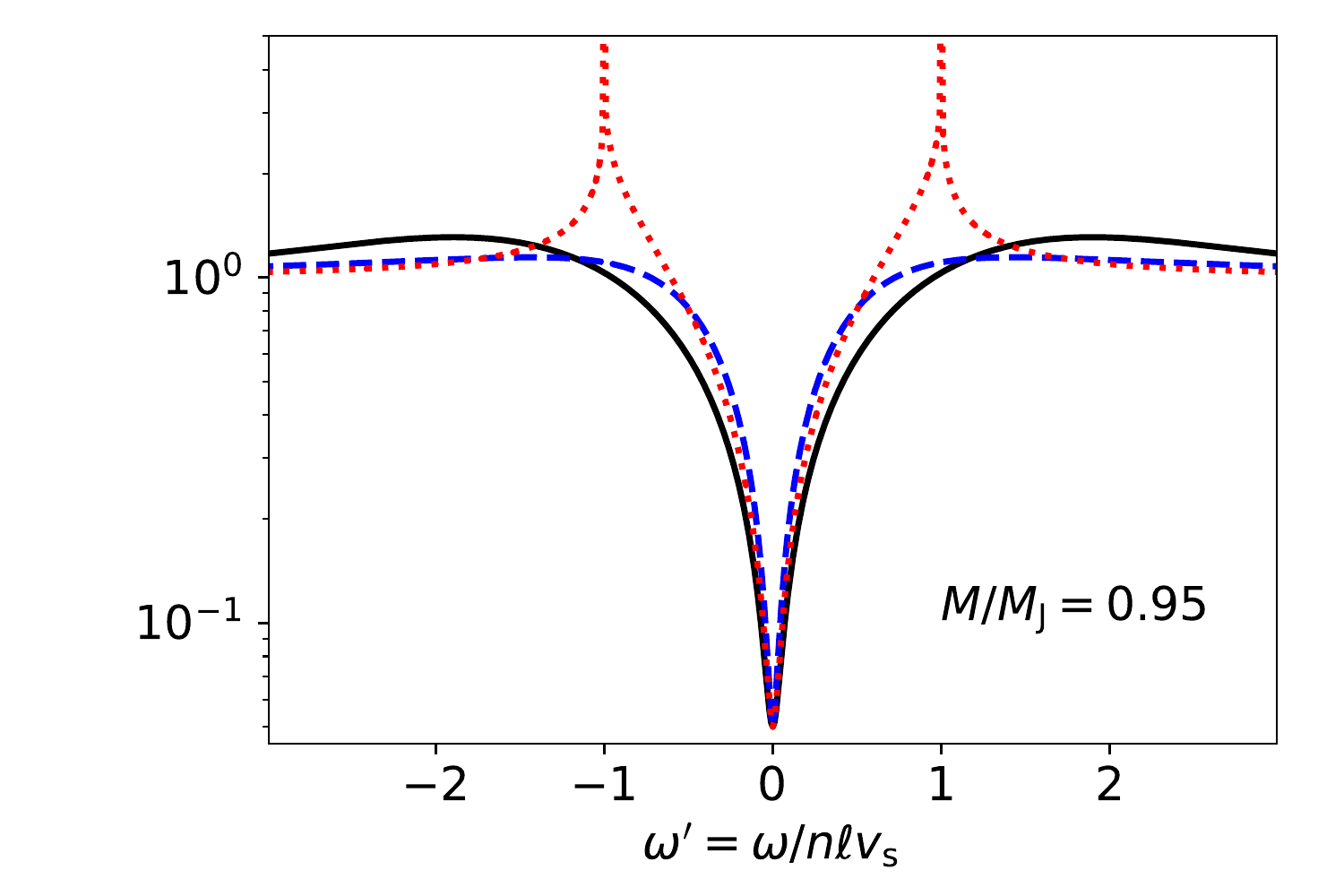}
  \caption{Modulus of the dielecric function 
    $\epsilon_\vn(\omega)\equiv1-\pol_\vn(\omega)$ for cube masses $0.8M_{\rm J}$ (left),
    $0.9M_{\rm J}$ (middle) and $0.95M_{\rm J}$ (right): the cube's
    response to an imposed wave of frequency~$\omega$ is proportional
    to  $(1-\pol_\vn(\omega))^{-1}$.  Here
    $|\vn|=1$.
  }
    \label{fig:logchi}
\end{figure*}

Figure~\ref{fig:logchi} shows how the dielectric function
$\epsilon_\vn(\omega)\equiv 1-\pol_\vn(\omega)$ of the $|\vn|=1$
response varies with the cube's mass for the Maxwellian, long-tailed
and isotropic top-hat DFs: from equations \eqref{eq:Phiresp}
and~\eqref{eq:In} the amplitude of the response to an imposed
$\Phi\stim_\vn=\e^{-\i\omega\tau}$ wave is proportional to
$1/\epsilon_\vn(\omega)$.  As the cube's mass approaches the
appropriate Jeans mass it responds more and more
vigorously to low-$\omega$ waves.  For fixed~$\omega$ the
polarizability~\eqref{eq:In} scales as $1/n^2$, which means that the
magnitude of the response shrinks rapidly as $|\vn|$ increases.

\section{Application to dynamical friction}

Now we turn to the problem of a cube being perturbed by a satellite of mass
$M\pert$ that is launched from $\vx=(0,0,0)$ with velocity $\vv\pert$.
Writing $\vJ\pert\equiv \ell\vv\pert$, the externally imposed
potential due to the satellite is
\begin{equation}
\Phi\stim_{\vn}(\tau)
=\frac{M\pert V_\vn}{(2\pi)^3\ell^3}\e^{-i\vn\cdot\vJ\pert\tau}.
\label{eq:stirpot}
\end{equation}
We first consider the case in which $\vv\pert$ and therefore
$\vJ\pert$ are held constant.  That is, we use the satellite to
``stir'' the cube.  Then in Section~\ref{sec:damping} below we release the satellite and
account for the back reaction of the cube's response.

An important point to note is that, to linear order, this stirring
imparts no net momentum to the cube's stars, nor does it heat them:
the momentum and energy of the cube's response are set by the velocity
moments of $f_{(0,0,0)}(\vJ)$, which, from the full
CBE~\eqref{eq:fullCBE}, can develop only through the nonlinear coupling
of $f_{\vm}$ with $\Phi_{-\vm}$.

\subsection{Steady-state linear response}
\label{sec:steadystatewake}
To simplify the problem we suppose that the stirred system settles
into a dynamical equilibrium, in which the response is related to the
stimulus~\eqref{eq:stirpot} by the time-independent relation
\begin{equation}
\Phi\resp_\vn=r_\vn\Phi\stim_{\vn},
\label{eq:linPhi}
\end{equation}
where the complex coefficient $r_\vn$ gives the amplitude and phase of
the $\vn^{\rm th}$ mode of the potential response with respect to the
stimulus.  We can find $r_\vn$ by substituting~\eqref{eq:linPhi}
into our expression~\eqref{eq:Phiresp} for $\Phi\resp_\vn(\tau)$ and
rearranging.  If we omit the self-gravity
of the response by taking $\Phi=\Phi\stim$ in the integrand
of~\eqref{eq:Phiresp}, then we obtain
\begin{equation}
  \Phi\resp_\vn=\pol_\vn(\vn\cdot\vJ\pert)\Phi\stim_\vn,
  \label{eq:notscresponsePhi}
\end{equation}
where $\pol_\vn(\omega)$ is the polarizability~\eqref{eq:In}.  This
$\pol_\vn(\vn\cdot\vJ\pert)\Phi_\vn\stim$ is the so-called ``undressed'' response, because
it does not account for the ``dressing'' that the response itself
contributes to the potential.  On the other hand, it is just as easy
to include the self gravity of the response by taking
$\Phi=\Phi\stim+\Phi\resp$ in the integrand of~\eqref{eq:Phiresp}.
Then, after some rearrangement, we obtain the ``dressed'' response
potential as
\begin{equation}
  \Phi\resp_\vn
  =\frac{\pol_\vn(\vn\cdot\vJ\pert)}{1-\pol_\vn(\vn\cdot\vJ\pert)}\Phi\stim_\vn.
  \label{eq:scresponsePhi}
\end{equation}
The electrostatic analogue of this equation is
$-\vP=(\pol/\epsilon)\vD$, where $\vP$ is the polarization vector, $\vD$
the electric displacement, and $\epsilon=1-\pol$ is the permittivity
or dielectric function.

We now have everything we need to calculate the steady-state dynamical
friction force in the following.
The galaxy's response
$\Phi\resp(\vtheta,\tau)=\sum_\vn \Phi\resp_\vn(\tau)\e^{\i\vn\cdot\vtheta}$
exerts a drag force $-M\pert\nabla\Phi\resp(\vtheta\pert)$ on the
satellite.  Using~\eqref{eq:notscresponsePhi}
and~\eqref{eq:scresponsePhi}, the corresponding deceleration is
\begin{equation}
  \begin{split}
    \dot\vJ\pert&=-\frac\partial{\partial\vtheta}\Phi\resp(\vtheta\pert)
    \\
    &=\i\sum_{\vn}\frac{ M\pert V_\vn}{(2\pi\ell)^3}\vn\times
    \begin{cases}
      \pol_\vn(\vn\cdot\vJ\pert), & \hbox{undressed,}\\
      \frac{\pol_\vn(\vn\cdot\vJ\pert)}{1-\pol_\vn(\vn\cdot\vJ\pert)}, &  \hbox{dressed}.
    \end{cases}
\end{split}
\label{eq:deceleration}
\end{equation}

For later
use (Section~\ref{sec:LB}) we note that the DF response $\tilde
f_\vn(\vk)$ can be obtained in the same way by
assuming that it behaves as
\begin{equation}
  \label{eq:linf}
  \frac{MV_\vn \tilde f_\vn(\vk)}{\ell^3}=r_\vn(\vk) \Phi\stim_\vn,
\end{equation}
which is the steady-state assumption~\eqref{eq:linPhi}
generalised to $\vk\ne\vzero$ by having a $\vk$-dependent response coefficient~$r_\vn(\vk)$.
Substituting this expression for $\tilde f_\vn(\vk)$ into equation~\eqref{eq:DFresplin} with
$\Phi=\Phi\stim$, the response DF~$\tilde f_\vn(\vk)$ in the undressed case is given by
\begin{equation}
  \frac{MV_\vn\tilde f_\vn(\vk)}{\ell^3}
  =
  \tilde\pol_\vn(\vn\cdot\vJ\pert,\vk)
  \Phi\stim_\vn,
\label{eq:nonscresponsef}
\end{equation}
while in the self-consistent, dressed case
($\Phi=\Phi\stim+\Phi\resp$) it is given by
\begin{equation}
  \frac{MV_\vn\tilde f_\vn(\vk)}{\ell^3}
  =
  \frac{\tilde\pol_\vn(\vn\cdot\vJ\pert,\vk)}{1-\pol_\vn(\vn\cdot\vJ\pert)}
  \Phi\stim_\vn,
\label{eq:scresponsef}
\end{equation}
where, in both cases, the $\vk$-dependent polarizability is
\begin{equation}
  \tilde\pol_\vn(\omega,\vk)\equiv -\frac{MV_\vn}\ell
    \int_0^\infty[\vn\cdot\vk+\vn^2\tau]
  \e^{\i\omega\tau}\tilde\fzero(\vk+\vn\tau)\,\d\tau.
\end{equation}

\begin{figure*}
  \centering
  \includegraphics[width=0.32\hsize]{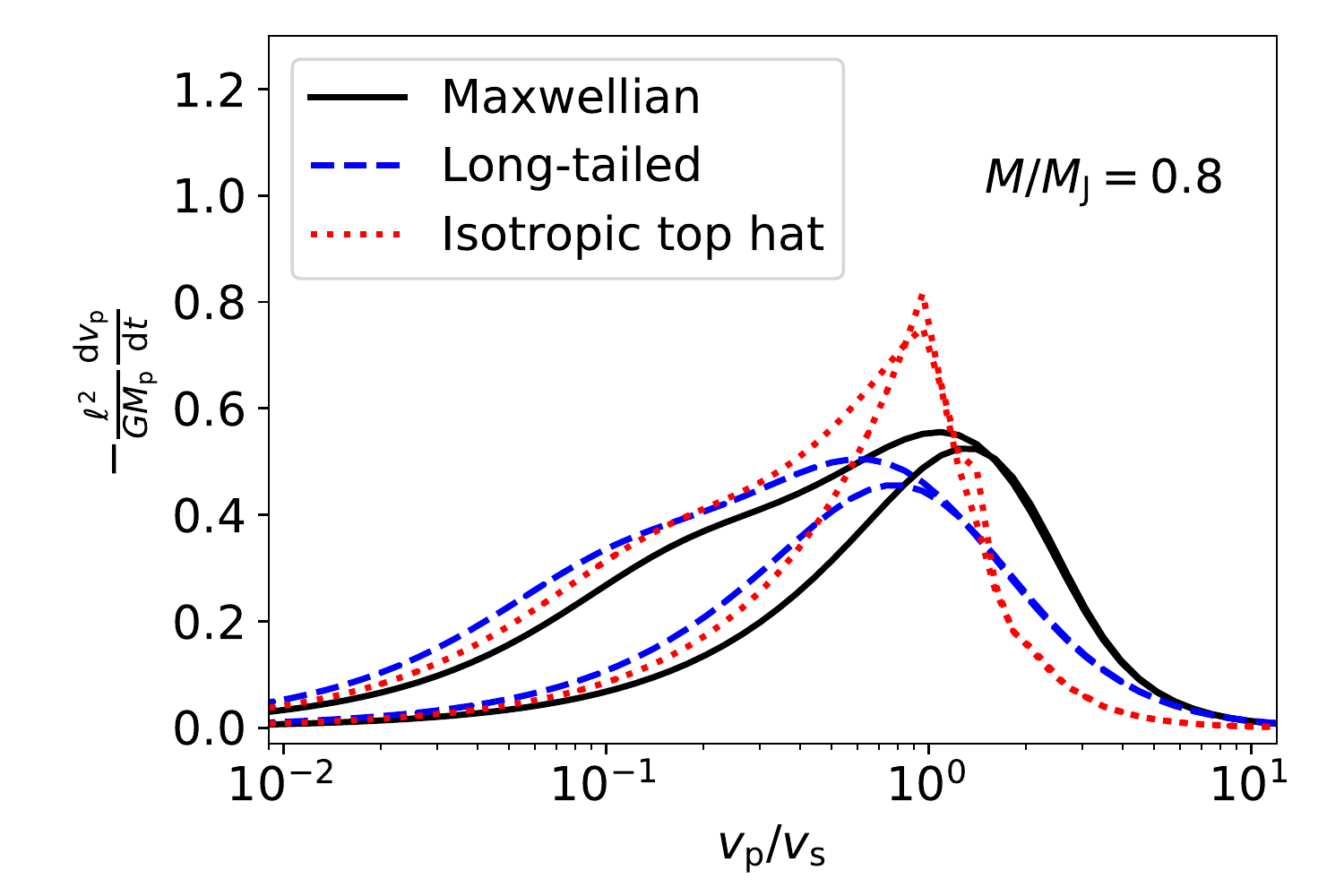}
  \includegraphics[width=0.32\hsize]{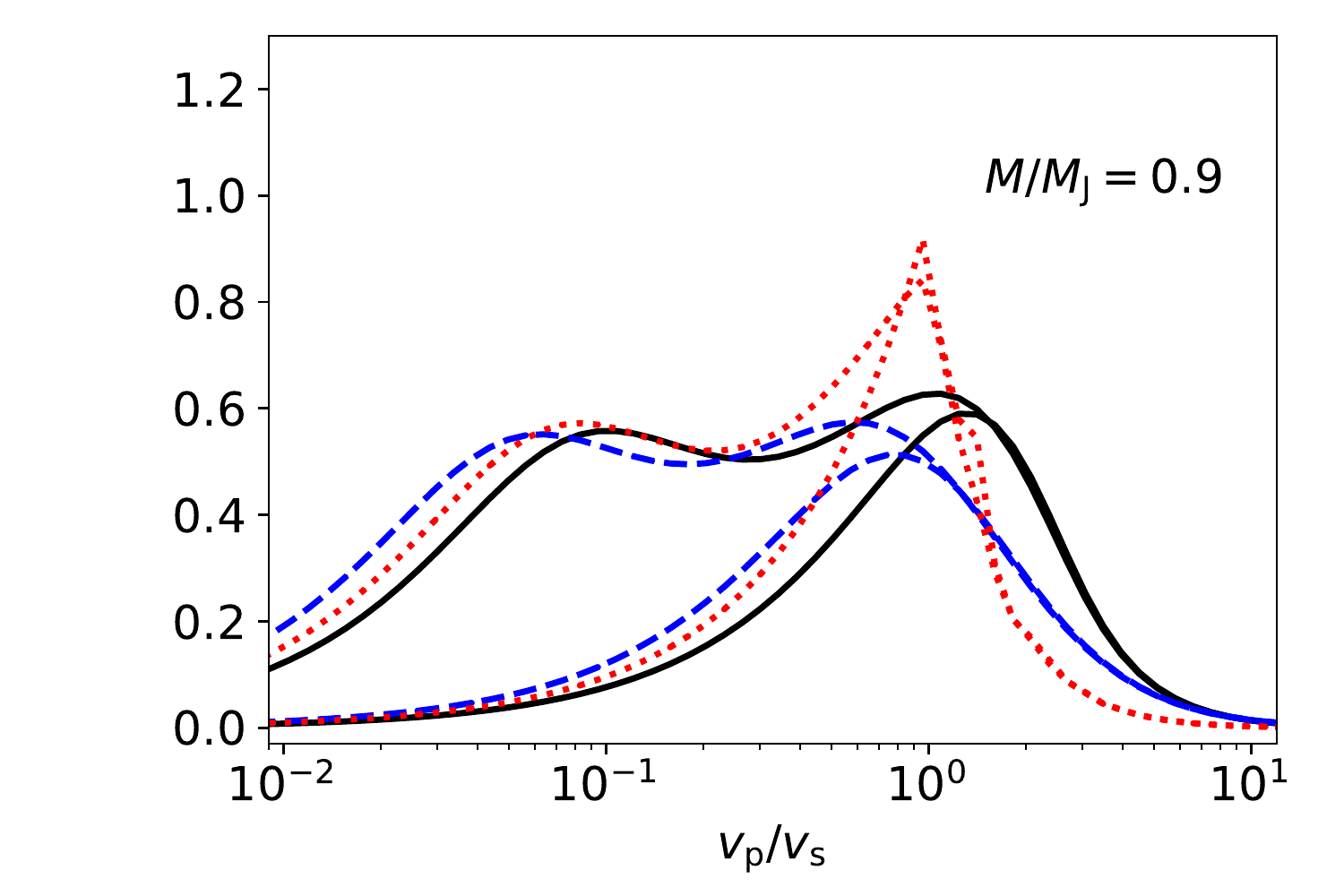}
  \includegraphics[width=0.32\hsize]{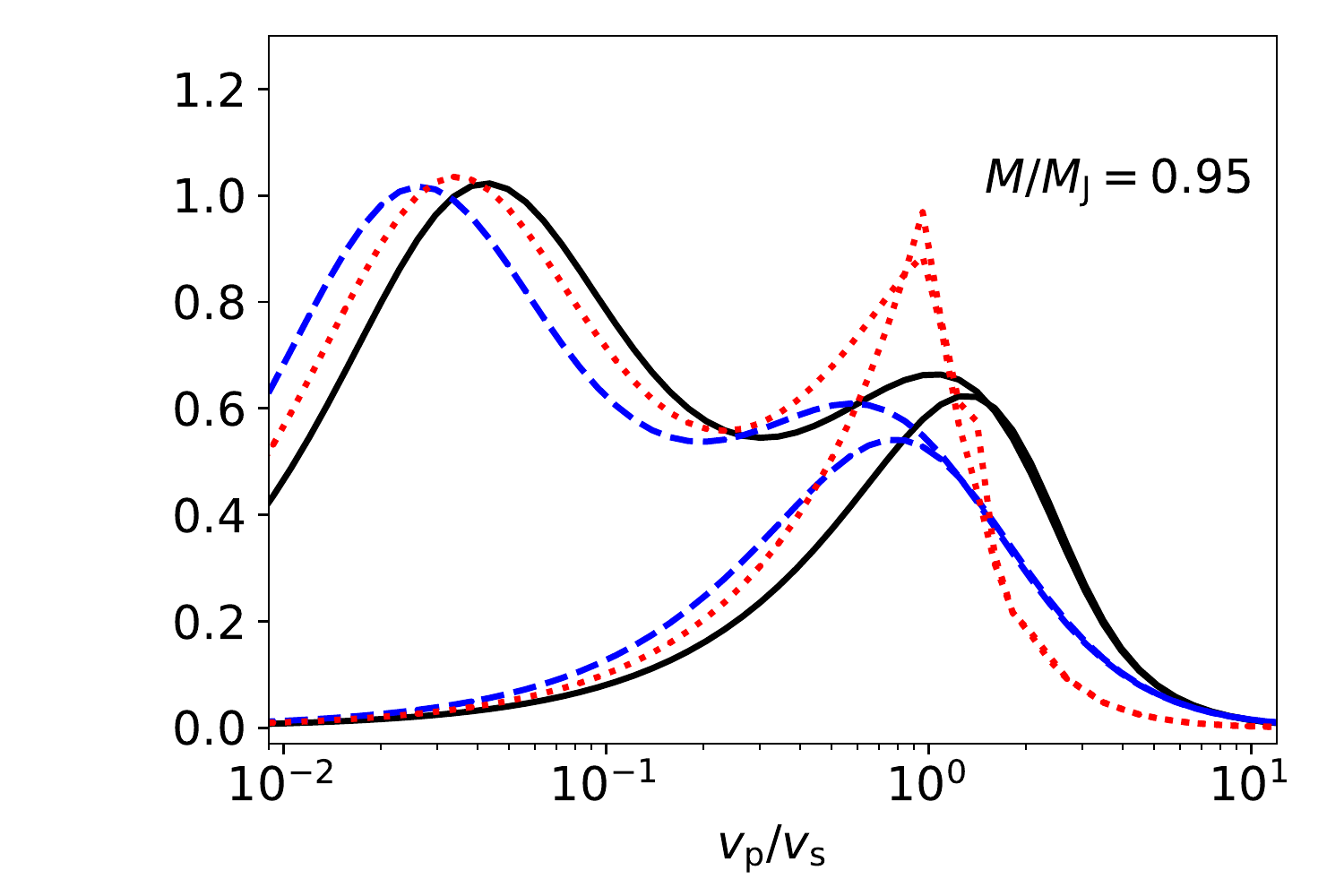}
  \caption{The deceleration from dynamical
    friction~\eqref{eq:deceleration} experienced when
    pulling a satellite of mass $M\pert$ through various cubes with
    constant speed~$v\pert$.  From left to right, the panels plot
    results for cubes of mass 0.8, 0.9 and 0.95 times the Jeans mass
    $M_{\rm J}$ (equations~\ref{eq:MaxwellMJ}, \ref{eq:IsoTopHatMJ}).
    Within each panel the pair of solid curves plot results for cubes
    with a Maxwellian DF, the upper curve showing the deceleration
    calculated from the {\it dressed} response of the cube to the
    satellite (that is, accounting for the self-gravity of the
    response), the lower from the {\it undressed} response (that is,
    ignoring the self gravity of the response).  The dashed and dotted
    curves plot corresponding
    results for the long-tailed and isotropic top-hat DFs,
    respectively.  }
  \label{fig:ssdrag}
\end{figure*}

\begin{figure}
  \includegraphics[width=0.45\hsize]{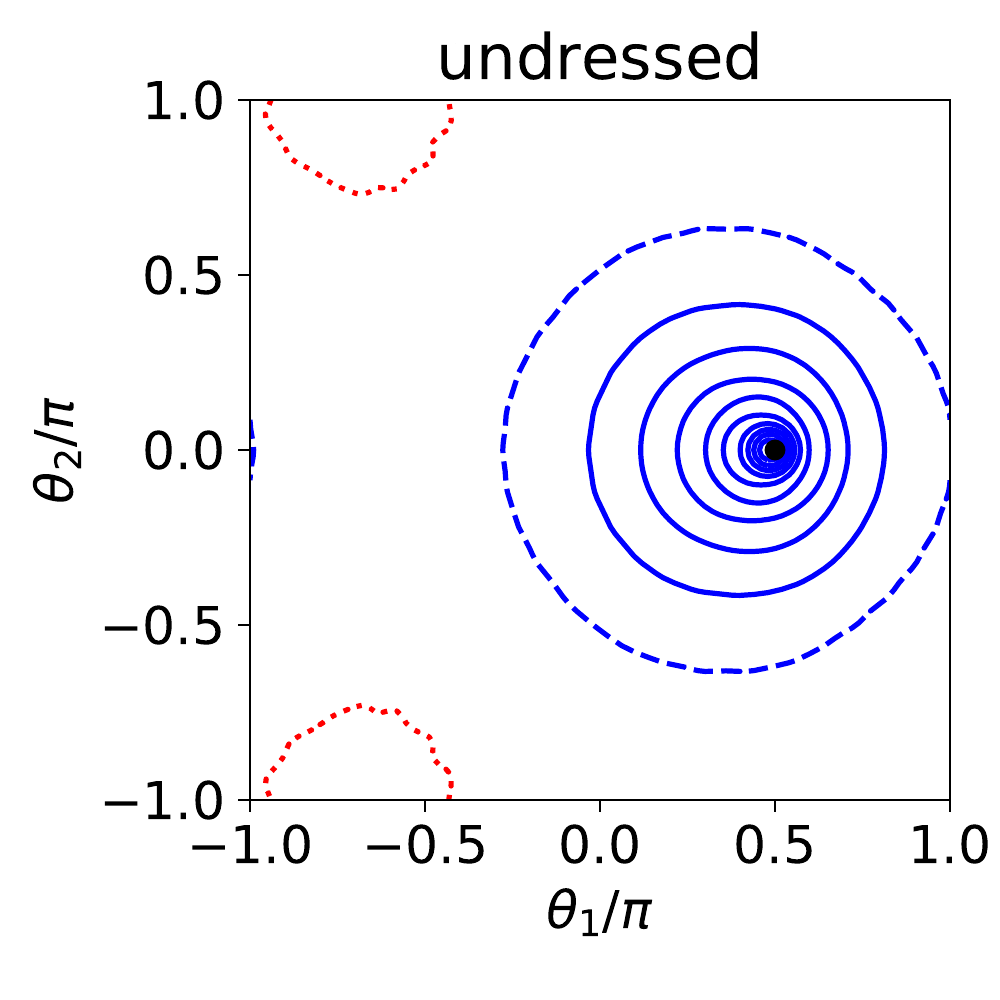}
  \includegraphics[width=0.45\hsize]{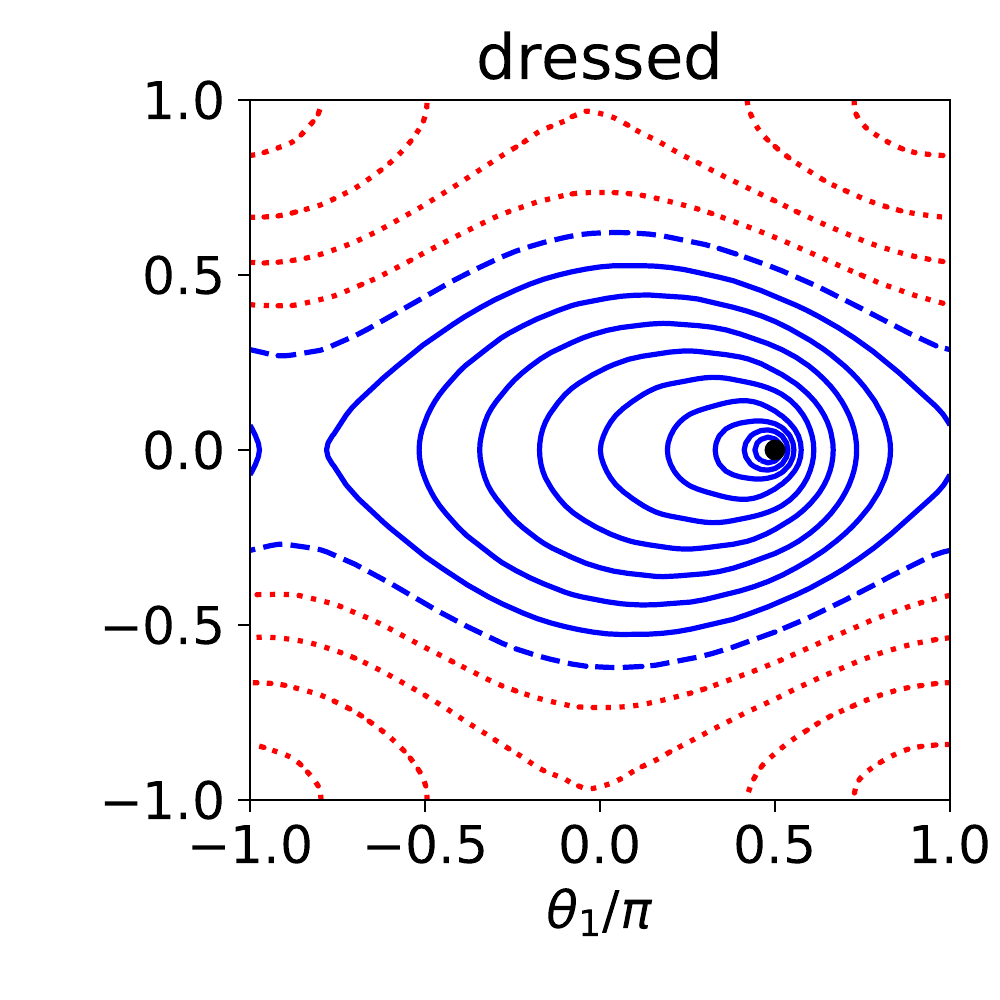}

  \caption{Response density, projected along the $\theta_3$ axis, for
    a satellite dragged with speed $J_1=0.3\vsc$ through a cube
    of mass $M=0.8M_{\rm J}$ with a Maxwellian DF.  The satellite's
    position is marked by the heavy black dot.  The panel on the
    left shows the undressed response, the panel on the right the dressed response.
    The same, linearly-spaced, contour levels are used in both
    panels.  Solid contours denote positive projected density,
    dotted contours negative density.
  }
  \label{fig:dragresp}
\end{figure}

\subsection{Chandrasekhar's formula}
\label{sec:chandra}
In the undressed case, the expression~\eqref{eq:deceleration} for the
drag simplifies if the velocity distribution within the cube is
isotropic (that is, if $\fzero(\vJ)=\fzero(|\vJ|)$).
Taking~$V_\vn$ from~\eqref{eq:Vn} and~$\pol_\vn$ from~\eqref{eq:In} we
have that 
\begin{equation}
\dot\vJ\pert=-\i\frac{(4\pi G)^2MM\pert}{(2\pi)^3}S,
\label{eq:Jdotnotsc}
\end{equation}
where $S$ is a sum over the contributions from different wavenumbers:
\begin{equation}
  S=\sum_\vn\frac{\vn}{n^2}\int_0^\infty\tau\e^{\i\vn\cdot\vJ\pert\tau}\tilde
  F(\vn\tau)\,\d\tau.
\end{equation}
Introducing spherical polar coordinates $(n,\theta,\phi)$ for $\vn$
with the polar axis parallel to $\vJ\pert$ and then approximating
the sum over~$\vn$ as an integral, we have
\begin{equation}
  \begin{split}
    S\simeq2\pi\log\left(\frac{n_{\rm max}}{n_{\rm min}}\right)\frac{\vJ\pert}{|\vJ\pert|}
    \int_0^\infty \d k  \int_{-1}^{+1}\d\mu\,\mu\e^{\i k\mu J\pert }
    k\tilde F(k),
  \end{split}
  \label{eq:Ssum}
\end{equation}
in which $\mu=\cos\theta$ and $k=n\tau$.  Similarly, expressing
equation~\eqref{eq:FT} in spherical polar coordinates gives
\begin{equation}
\tilde F(k)=2\pi\int_0^\infty\d J\,
    J^2\fzero(J)\int_{-1}^{+1}\e^{-\i kJ\mu'}\,\d\mu'.
\end{equation}
Substituting this $\tilde F(k)$ into~\eqref{eq:Ssum}, performing the
integral over~$k$, then $\mu'$ and finally $\mu$, our
expression~\eqref{eq:Jdotnotsc} for the frictional deceleration in the
undressed case becomes
\begin{equation}
  \dot\vJ\pert = -
  (4\pi G)^2MM\pert\log\left(\frac{n_{\rm max}}{n_{\rm min}}\right)
  \frac{\vJ\pert}{J\pert^3}\int_0^{J\pert}
  F(J)J^2\,\d J.
\label{eq:chandra}
\end{equation}
This is identical to the usual Chandrasekhar dynamical friction
formula
\citep[e.g.,][]{BinneyGalacticDynamicsSecond2008,ChavanisKinetictheoryspatially2013b}
with $\log(n_{\rm max}/n_{\rm min})$ playing the role of the Coloumb
logarithm, $\log\Lambda$.  An important difference, however, is that
the usual derivation of the formula assumes local two-body encounters
and integrates them exactly, which removes the small-scale cutoff.  In
our derivation, the unperturbed orbits are straight lines, not
two-body hyperbolae; to obtain the inner cutoff correctly we would
need to include the $O(\epsilon^2)$ nonlinear response.

\subsection{Examples of the steady-state response}

Figure~\ref{fig:ssdrag} plots how the drag from dynamical
friction~\eqref{eq:deceleration} depends on the perturber's speed
$v\pert=J\pert/\ell$ and on the mass of the cube, $M/M_{\rm J}$
for each of the stable DFs listed in Table~\ref{tab:fzeros}.  For
these curves we have taken $n_{\rm min}=1$ and
$n_{\rm max}=18\simeq GM/\ell\vsc^2$.  Figure~\ref{fig:dragresp}
shows a typical example of the density response from which this drag
is computed, demonstrating that the dressed response is much stronger
than the undressed one.

Although we do not plot it here, the drag computed from the undressed
response agrees very well with the Chandrasekhar dynamical friction
formula~\eqref{eq:chandra}.  The only notable difference is that the
Chandrasekhar formula predicts a sharp cutoff in the drag when
$v\pert>\vsc$ in the isotropic top-hat model: according to it stars
that move faster than the satellite should not contribute to the
frictional force.  Figure~\ref{fig:ssdrag} shows that in our sum over
discrete spatial modes~$\vn$ this cutoff for the isotropic top-hat
model is softened slightly, with further signs of the discreteness
occuring when $v\pert$ is an integer multiple of~$\vsc$.

In the undressed models the frictional deceleration scales linearly with
both $M$ and~$M\pert$.  When dressing is included, however, as
$M\to M_{\rm J}$ the system becomes increasingly responsive to the
perturbing satellite: instead of remaining close to 1 everywhere, the
magnitude of the dielectric function $1-\pol_\vn(\vn\cdot\vJ\pert)$ can
approach 0, leading to an arbitrarily large response
\eqref{eq:scresponsePhi}.  This is particularly
noticeable for slowly moving satellites with $v\pert \ll\vsc$.
The right-hand panel of Figure~\ref{fig:ssdrag} shows that the
undressed calculation can underestimate the friction by orders of
magnitude when $v\pert\sim0.1\vsc$ for all three DFs.  For a
satellite moving with $\vv\pert=(0,0,v\pert)$ the
dressing predominantly affects the $\vn=(0,0,\pm1)$ contributions
to the response.  The contribution to the other modes is weaker than
the undressed response.

We note that the deceleration due to dynamical friction depends mostly
on $M/M_{\rm J}$, on the speed of the satellite compared to typical
stars $v\pert/\vsc$, and on whether or not one includes dressing.  The
shape of the cube's $F(\vv)$ distribution is a smaller effect, the
most noticeable difference among the three DFs being in the detailed
behaviour of the deceleration around $v\pert/\vsc\simeq1$ in the
isotropic top-hat model.

\begin{figure*}
  \includegraphics[width=0.45\hsize]{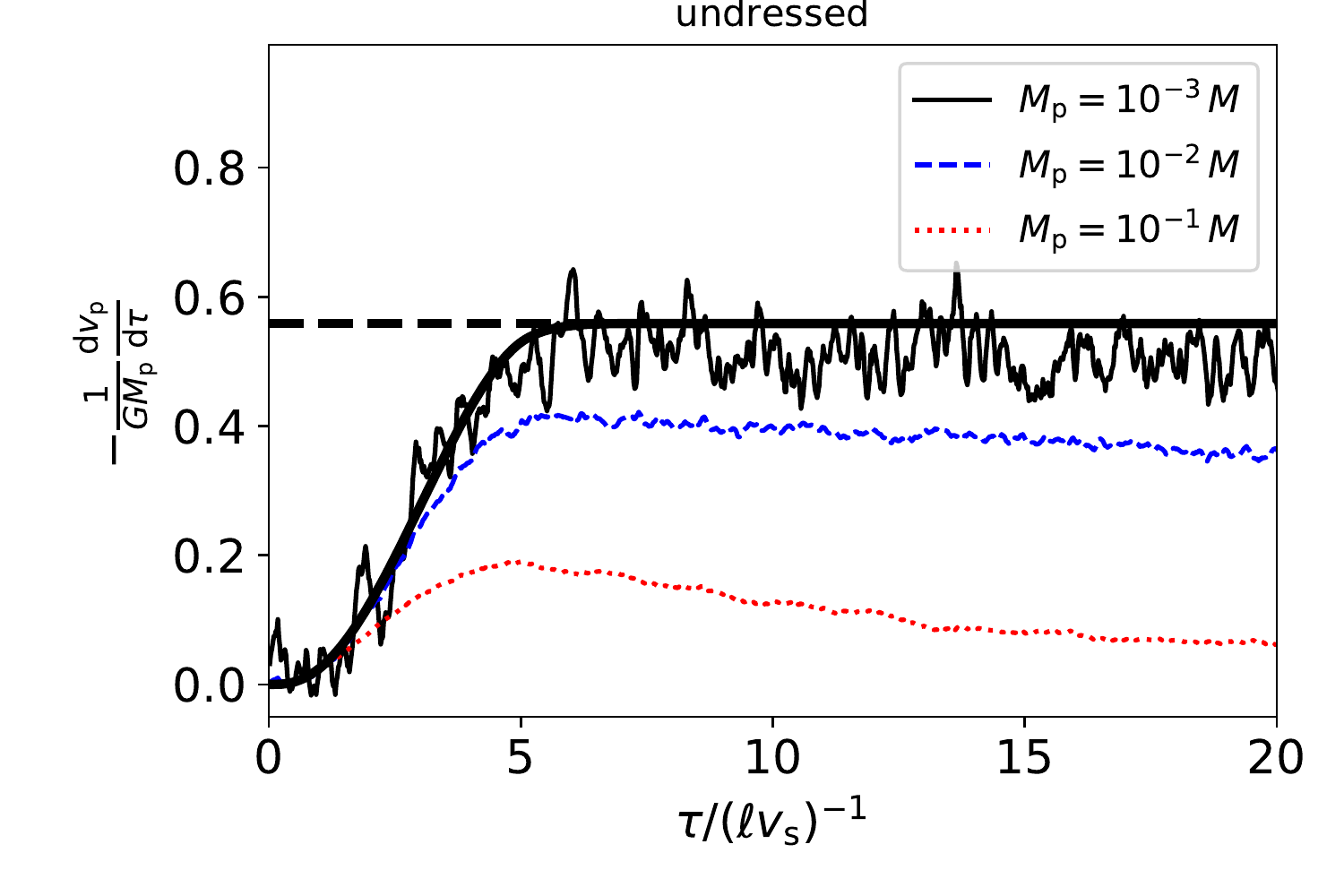}
  \includegraphics[width=0.45\hsize]{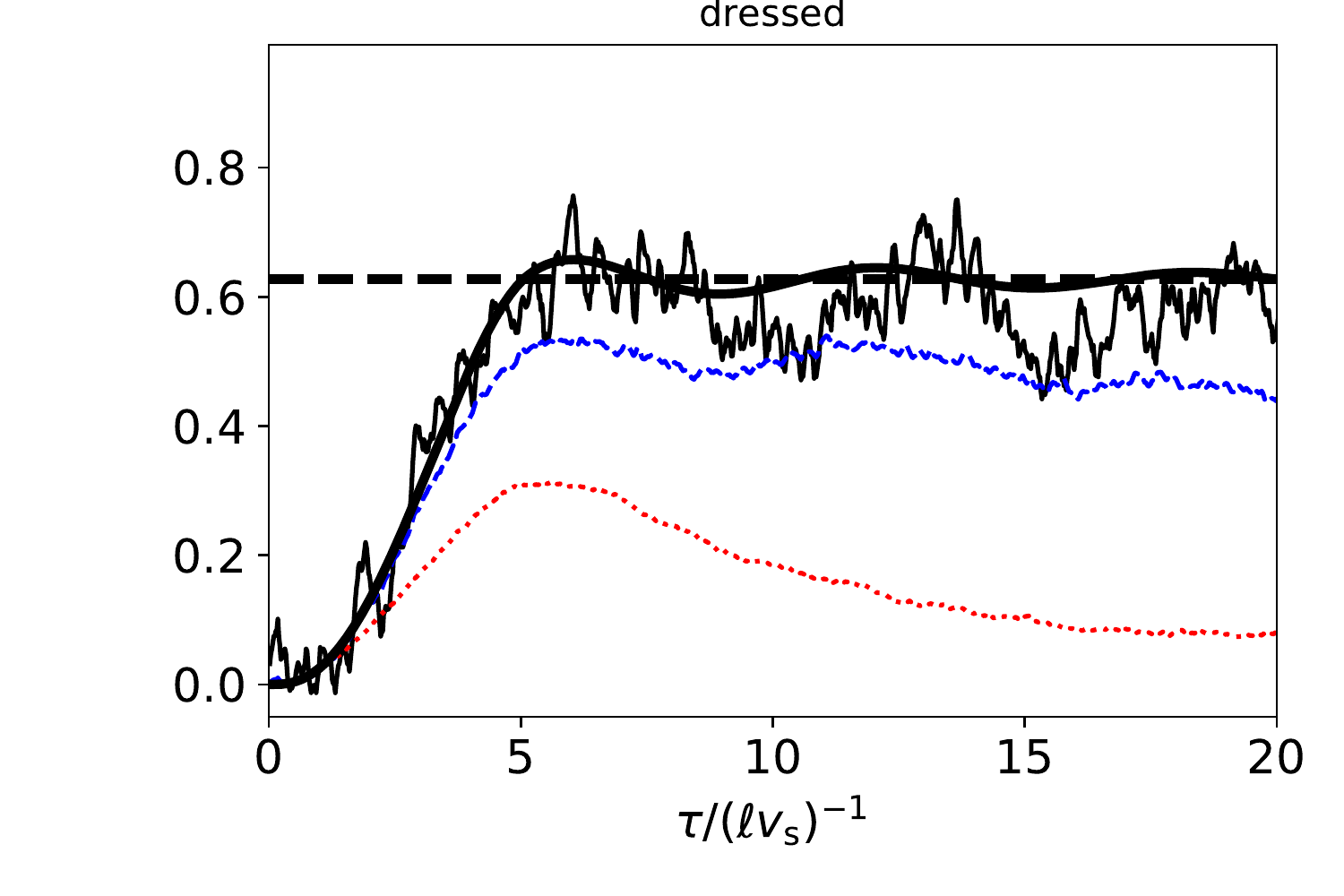}
  \caption{Time evolution of the friction experienced by a satellite
    that is pulled with constant speed $V_{\rm p}=\vsc$ through a 
    cube of mass $M=0.9M_{\rm J}$ with a Maxwellian DF.
    The satellite's mass grows with time according
    to~\eqref{eq:Mt} over $t=5$ time units.
    The left and right panels show the drag force that the cube's undressed 
    and dressed response, respectively, exert on the satellite, normalized
    by the square of the satellite's final mass.
    The heavy solid curves plot results obtained by solving the
    linear Volterra equation~\eqref{eq:Phiresp}.
    At late times these are in excellent agreement with the steady-state
    prediction~\eqref{eq:deceleration} (horizontal dashed line).
    For comparison, the jagged curves plot
    the drag measured from $N$-body simulations of the cube's
    nonlinear response for satellites that grow to masses
    $M_{\rm p}=10^{-3}\,M$, $10^{-2}\,M$ and $10^{-1}\,M$.
    The simulations use $N=10^8$ particles to sample the the cube's DF.}
  \label{fig:volterradrag}
\end{figure*}

\subsection{Time-dependent response: stirring}
\label{sec:volterrat}

In obtaining the results above we have assumed
(equation~\ref{eq:linPhi}) that a new equilibrium DF is established
within the cube: that is, that $f(\vx,\vv)$ is constant in a frame that moves with the
satellite's constant stirring speed $\vv\pert$.  The most immediate
way of testing this assumption is by using equation~\eqref{eq:Phiresp}
to obtain the time-dependent response of the cube to a satellite that
is introduced from time $\tau=0$.  

The satellite's influence enters through the $\Phi_\vn(\tau')$ factor in the
integrand of~\eqref{eq:Phiresp}.  We introduce the satellite slowly, its mass
growing as
\begin{equation}
  \begin{split}
    M\pert(\tau)&=M_{\rm p,final}\,\mu\left(\frac \tau{\tau_{\rm grow}}\right),\\
  \end{split}
\label{eq:Mt}
\end{equation}
with $\tau_{\rm grow}=5/\ell\vsc$ and
\begin{equation}
  \mu(s)=
  \begin{cases}
    0, & s<0,\\
    3s^2-2s^3, &0\le s\le1,\\
    1, &s>1.
  \end{cases}
\end{equation}
To solve~\eqref{eq:Phiresp} we use the
trapezoidal rule with step $\Delta\tau=0.01/\ell\vsc$ to approximate the
integral on the right-hand side as a sum over the values of
$\Phi_\vn$ at discrete times $\tau=0$, $\Delta\tau$, ...., $j\Delta\tau$.
Rearranging this gives an expression for $\Phi_\vn\resp(j\Delta\tau)$
as a sum over the values of $\Phi_\vn$ at earlier times $\tau=0$, $\Delta\tau$, ...,
$(j-1)\Delta\tau$ and of $\Phi_\vn\stim$ at time $j\Delta\tau$.

The heavy solid curve in Figure~\ref{fig:volterradrag} shows a typical
example of how the friction experienced by the satellite develops over
time according to this linear calculation.  The introduction of the
satellite excites some oscillations in the cube.  Once these have
damped, the long-term response is in excellent agreement with the
steady-state prediction from~\eqref{eq:deceleration}.  The larger
$M/M_{\rm J}$ is, the longer this initial ringing takes to die away.

The same Figure plots the friction measured from $N$-body simulations that
include the full nonlinear response of the cube to the satellite.
The simulation method is described in
Appendix~\ref{sec:Nbody}.  It is clear from the
Figure that the linear response calculation described above is not
a particularly good approximation to the full nonlinear response for any
$M\pert\lesssim 10^{-2}M$.  The most obvious shortcoming of the linear
calculation is that it ignores the momentum transferred to the cube's
stars as the satellite is dragged through the system.  In the $N$-body
simulations this leads to a gradual decrease in the dynamical
friction force as the cube's stars speed up to match the satellite, a
phenomenon that is not captured by the linear calculation.

\begin{figure*}
  \includegraphics[width=0.45\hsize]{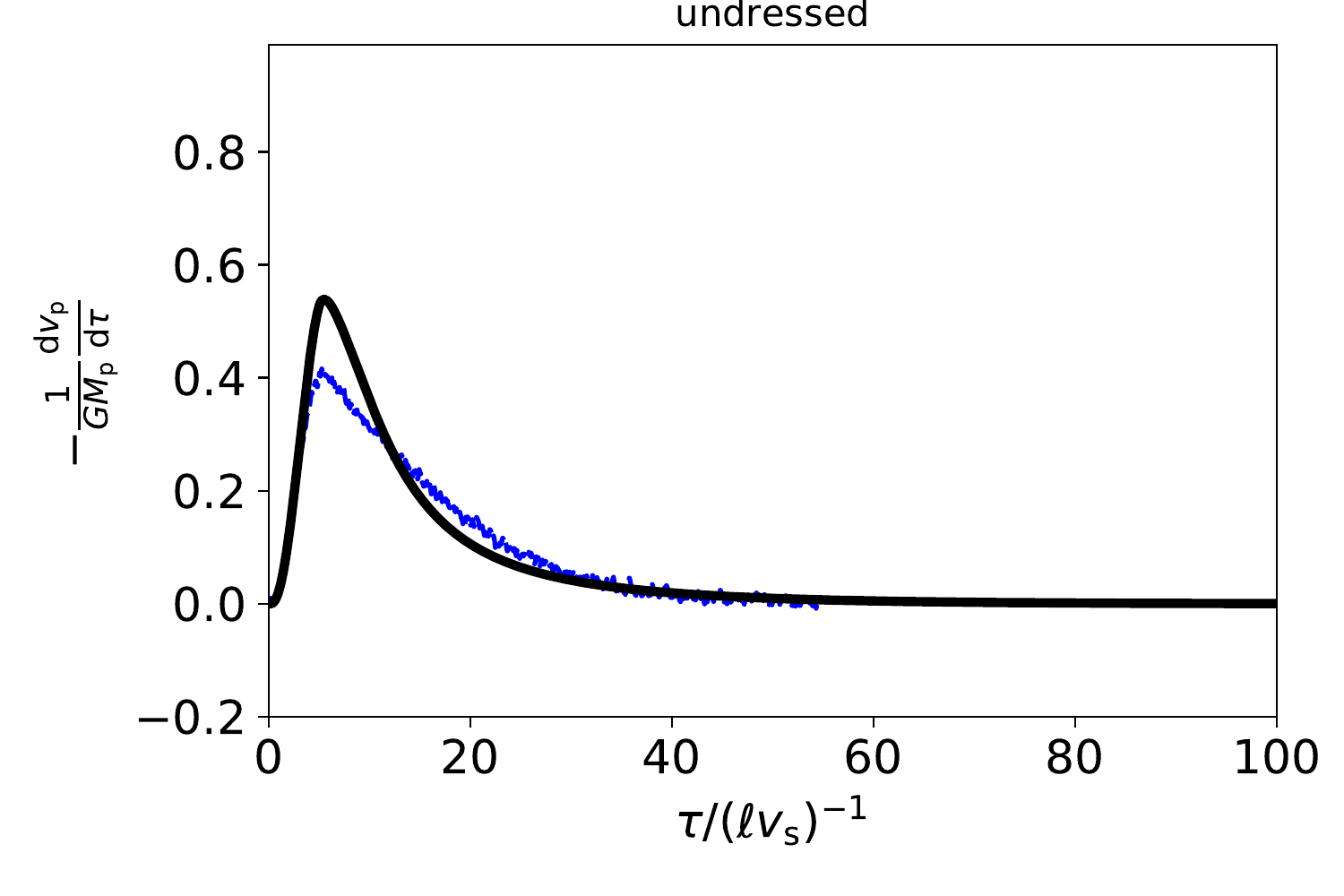}
  \includegraphics[width=0.45\hsize]{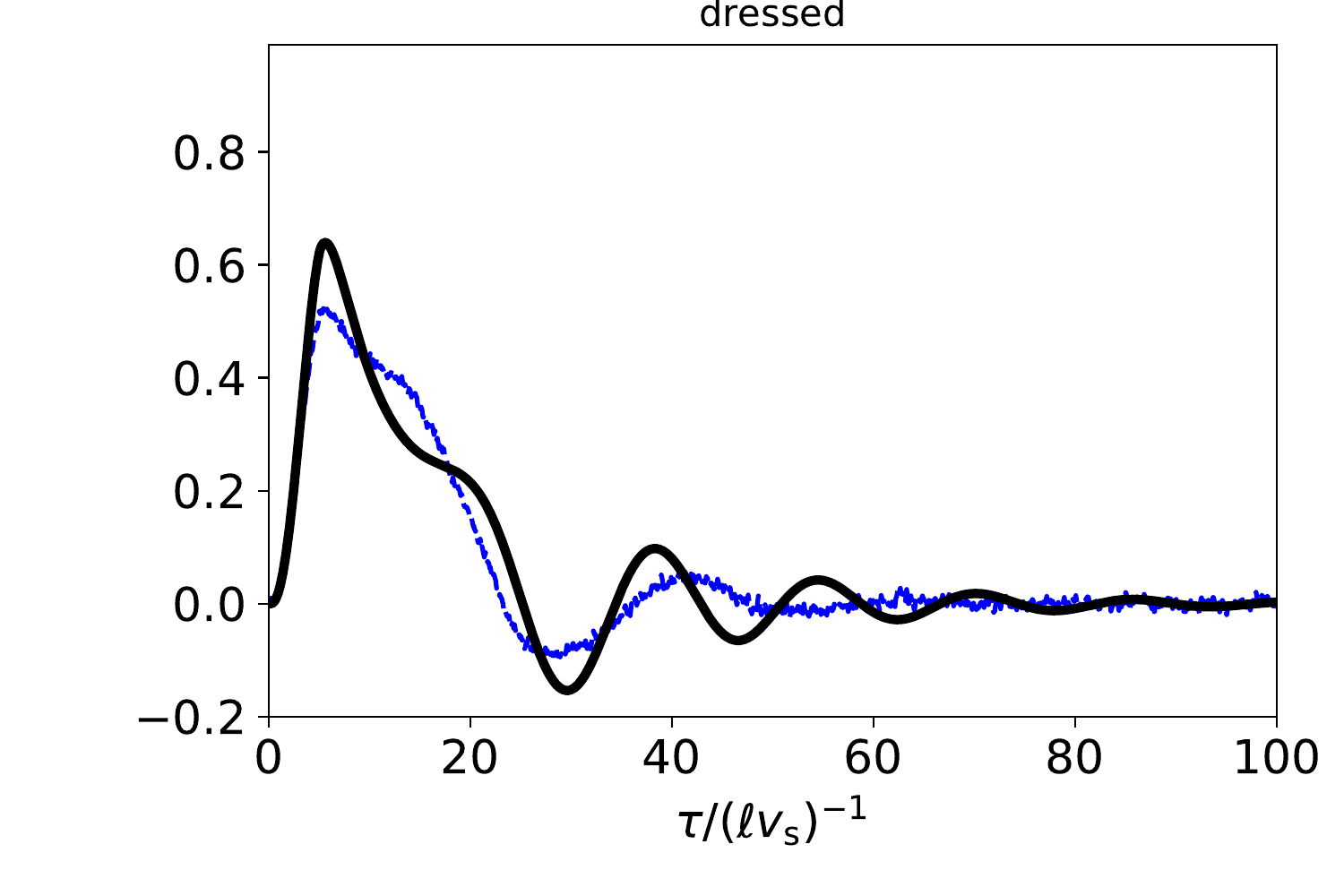}

    \includegraphics[width=0.45\hsize]{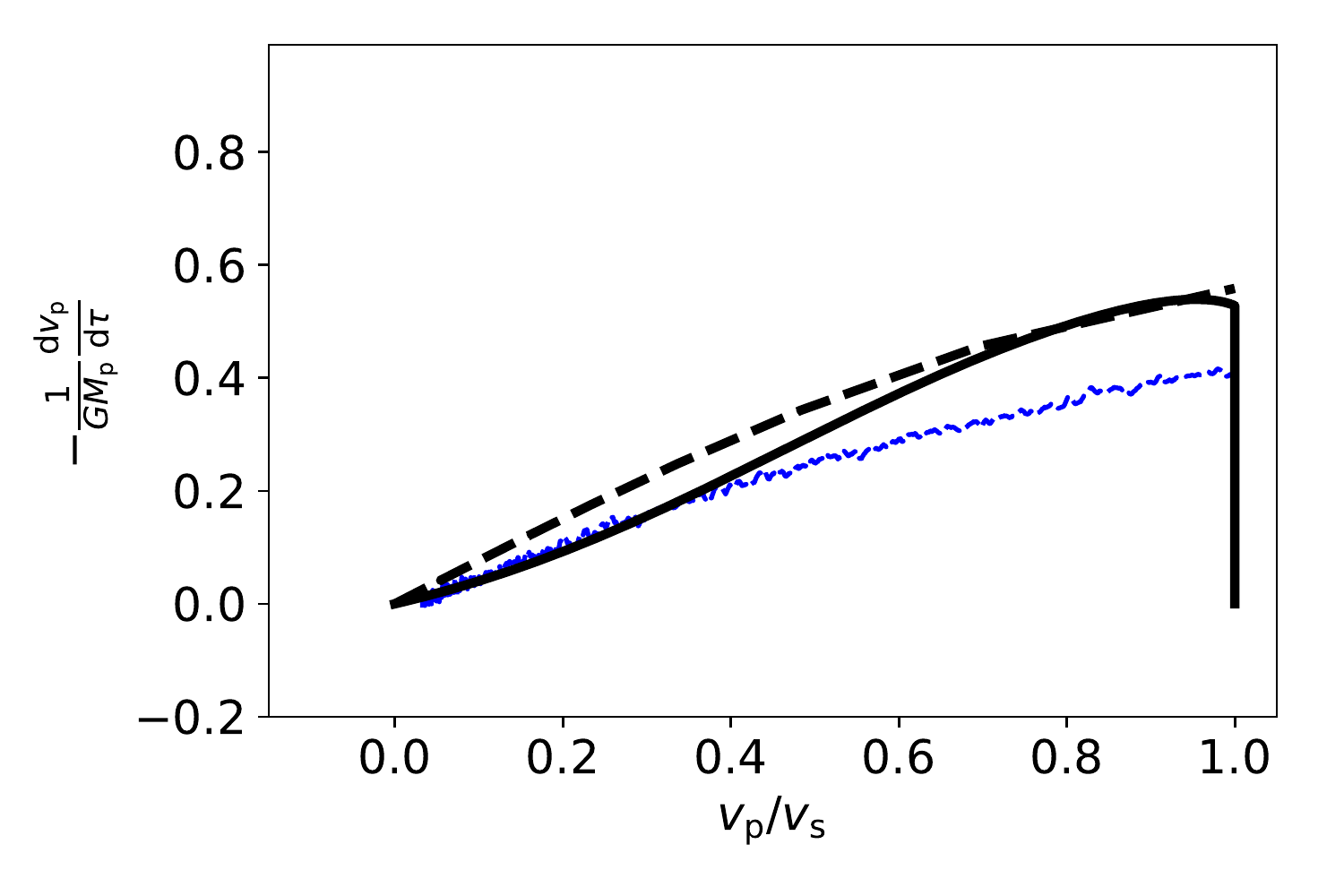}
  \includegraphics[width=0.45\hsize]{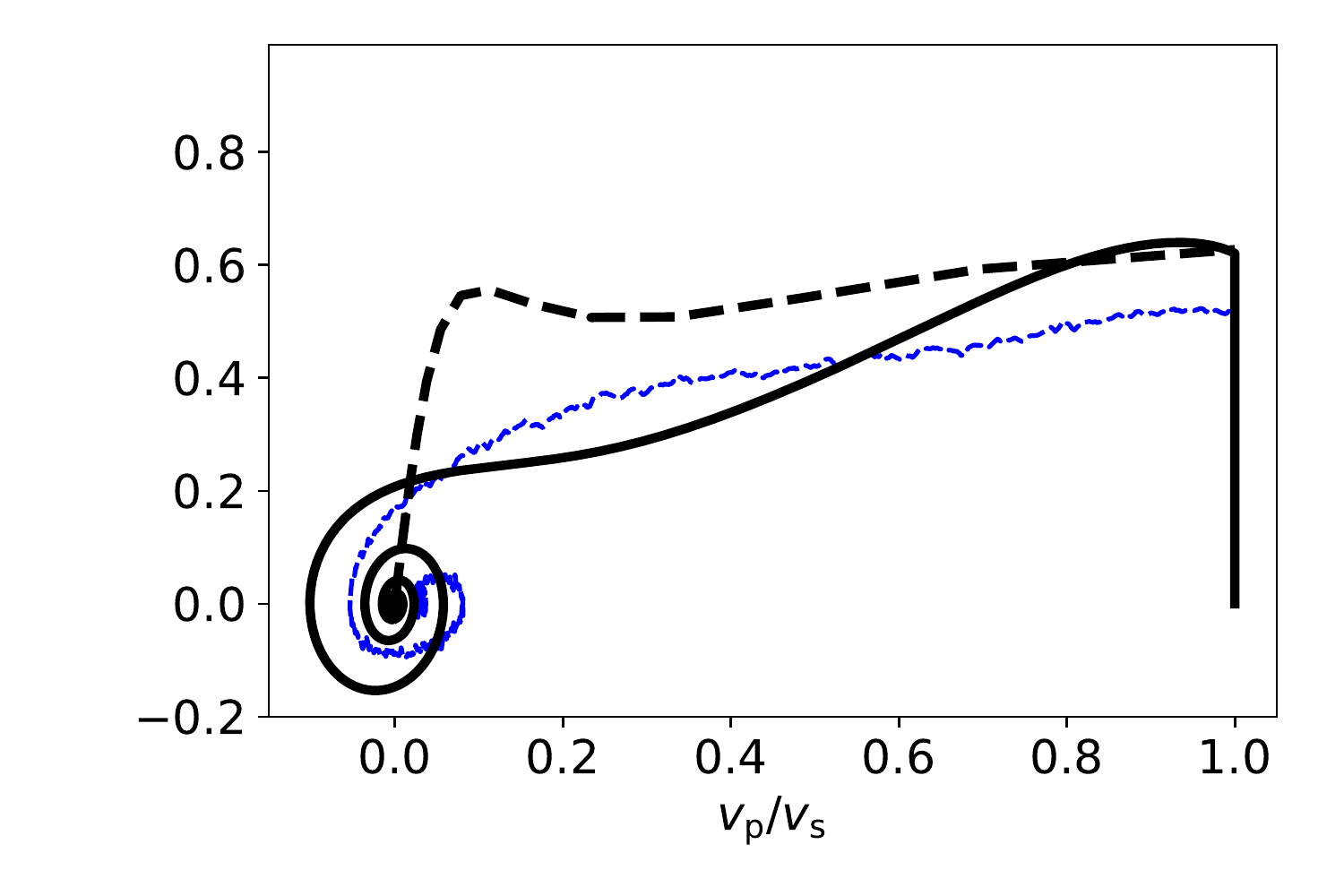}

  \caption{Damping of the motion of a satellite after it has been dragged
    through a cube of mass $\mass=0.9\,M_{\rm J}$ with a Maxwellian
    DF.  As in Figure~\ref{fig:volterradrag}, the satellite's mass
    grows from 0 to $10^{-2}\,M$ between $t=0$ and $t=5$, during which
    time it is dragged through the cube with constant speed
    $v\pert/\vsc=1$.
    At $t=5$ it is released, its subsequent motion governed by the
    cube's response.  The left and right panels show results when
    this response is undressed and dressed, respectively
    The top row shows the time evolution of the drag deceleration, the heavy solid
    curves giving the predictions of  time-dependent linear theory
    (Section~\ref{sec:damping}), the lighter dashed curves measurements
    from $N$-body simulation.
    The bottom row shows the same results, but plotted against the
    satellite's instantaneous speed $v\pert$: in this row the
    satellite starts at $(1,0)$, moves up as it is being dragged, then
    to the left when released.  For comparison the heavy dashed curve
    plots the deceleration predicted from the ``steady-state'' formalism of
    Section~\ref{sec:steadystatewake} using only the instantaneous
    value of $v\pert$.  }
    \label{fig:damp}
\end{figure*}

\subsection{Time-dependent response: damping}

\label{sec:damping}
In the preceding calculations we have supposed that the satellite is
pulled through the cube with constant velocity~$\vv\pert$.  What
would happen if we were to release the satellite?  To investigate this
we stir the cube by dragging the satellite through it to build
up a wake as before.  Then at some $\tau=\tau_{\rm release}$ we let go of the
satellite  and follow the subsequent evolution of the
coupled satellite-plus-wake system.  We use the 
following leapfrog integrator for the satellite's motion at times
$\tau\ge\tau_{\rm release}$:
\begin{equation}
  \begin{split}
    {\vtheta\pert}\left(\tau+{\scriptstyle\frac12}\Delta\tau\right)&={\vtheta\pert}(\tau)+{\scriptstyle\frac12}{\vJ\pert}(\tau)\Delta\tau,\\
    \vJ\pert(\tau+\Delta\tau)&=\vJ\pert-\i\sum_\vn\vn\Phi\resp_\vn\left(\tau+{\scriptstyle\frac12}\Delta\tau\right)\e^{\i\vn\cdot\vtheta\pert(\tau+{\scriptscriptstyle\frac12}\Delta\tau)}\Delta\tau,\\
    {\vtheta\pert}\left(\tau+\Delta\tau\right)&={\vtheta\pert}\left(\tau+{\scriptstyle\frac12}\Delta\tau\right)+{\scriptstyle\frac12}{\vJ\pert}(\tau)\Delta\tau.\\
  \end{split}
\label{eq:dampleap}
\end{equation}
The potential $\Phi_\vn\resp(\tau+\frac12\Delta\tau)$ in the middle,
``kick'', step is calculated as described in
Section~\ref{sec:volterrat} above, taking $\Phi\stim_\vn(\tau')$ from
the $\vtheta\pert(\tau')$ obtained from the
new satellite position~$\vtheta\pert(\tau+{\scriptstyle\frac12}\Delta\tau)$.

Figure~\ref{fig:damp} shows how the frictional deceleration evolves
for a satellite of mass
$10^{-2}\,M_{\rm J}$ that is released with $v\pert/\vsc=1$
within a Maxwellian cube of mass $M=0.9\,M_{\rm J}$.  If one considers only the
undressed response then this evolution is simple: the satellite slows
down until it comes to rest.  The dressed response is more
interesting, however: as the satellite slows down, the response that
is has induced in the cube's stars overtakes it, causing it to speed
up again, overtaking the response, leading to damped oscillations in
$v\pert(\tau)$.  The dashed curve in the lower right panel of
Figure~\ref{fig:damp} shows that the steady-state approximation used
in Section~\ref{sec:steadystatewake} completely fails to capture this
evolution.

Qualitatively the same phenomena are observed in $N$-body simulations
of the same system, the most notable difference being how the $N$-body
simulations exhibit smaller-amplitude oscillations and converge to a
final $v\pert\ne0$ that arises from the momentum imparted to the
cube's stars before the satellite is released.

\section{van Kampen modes}

In section~\ref{sec:linstab} we derived the dispersion
relation~\eqref{eq:dispersionreln}, the solutions to which are the
frequencies $\omega$ of the self-sustaining waves that the cube admits.
In general these frequencies are complex, which means that these waves
grow exponentially in either the forward or backward time direction.
These waves -- the so-called Landau modes -- are not the only waves
admitted by the cube, however.

Consider a self-consistent oscillation of an isolated cube
($\Phi\stim=0$) with real frequency~$\omega$:
\begin{equation}
 \begin{split}
   \tilde f_\vn(\vk,\tau|\omega)
   &
   =\tilde f_{\vn}(\vk|\omega)\e^{-\i\omega\tau},
\\
\Phi_\vn(\tau|\omega)
&
=\tilde \Phi_{\vn}(\omega)\e^{-\i\omega\tau},
  \end{split}
\end{equation}
in which, by Poisson's equation~\eqref{eq:poissonk}, 
$\tilde\Phi_{\vn}(\omega)=(MV_\vn/\ell^3)\tilde
f_{\vn}(\vzero|\omega)$.
Let $k_\parallel$ be the component of $\vk$ that is parallel to
$\vn$.  Then $\vn\cdot\vk=nk_\parallel$ and the linearized
Fourier-transformed CBE~\eqref{eq:kCBE} for $\tilde f_\vn(\vk_\perp,k_\parallel|\omega)$
becomes
\begin{equation}
  \label{eq:modef}
  \i\omega\tilde f_{\vn}(\vk_\perp,k_\parallel|\omega)
  +n \frac{\partial \tilde f_{\vn}}{\partial k_\parallel}(\vk_\perp,k_\parallel|\omega)
  -nk_\parallel\,\ell^2\tilde F(\vk_\perp,k_\parallel)\tilde\Phi_{\vn}(\omega)=0.
\end{equation}
The solutions are
\begin{equation}
  \begin{split}
&    \tilde f_{\vn}(\vk_\perp,k_\parallel|\omega)
    =\e^{-\i\omega k_\parallel/n}\times\\
&\quad  \left[
  \tilde f^\perp_{\vn}(\vk_\perp|\omega)
  +\tilde\Phi_{\vn}(\omega)\ell^2\int_0^{k_\parallel}\e^{\i\omega
      k_\parallel'/n}k_\parallel'\tilde F(\vk_\perp,k_\parallel')\,\d k_\parallel'
  \right],
\end{split}
\label{eq:vanKFT}
\end{equation}
in which the only constraint on the constant of integration
$\tilde f^\perp_{\vn}(\vk_\perp|\omega)$ is that
$\tilde f^\perp_{\vn}(\vzero|\omega)=\tilde f_{\vn}(\vzero|\omega)$.

These are the (Fourier-transformed)
\cite{vanKampentheorystationarywaves1955} modes of the cube.
To show
this explicitly, notice that
$k_\parallel \tilde F(\vk_\perp,k'_\parallel)$ is the Fourier
transform of $\partial F/\partial k'_\parallel$.  Then performing the
integral over $k_\parallel'$ gives
\begin{equation}
  \begin{split}
&    \tilde f_{\vn}(\vk_\perp,k_\parallel|\omega)
  =\e^{-\i\omega k_\parallel/n}
\tilde f^\perp_{\vn}(\vk_\perp|\omega)
\\
&\quad+
\tilde\Phi_{\vn}(\omega)\ell^2\e^{-\i\omega k_\parallel/n}
  \int\d^3\vJ'
  \frac{1-\e^{\i k_\parallel(\omega-\vn\cdot\vJ')/n}}{\omega-\vn\cdot\vJ'}
  \vn\cdot\frac{\partial F}{\partial \vJ'}.
\end{split}
\label{eq:vanKFT2}
\end{equation}
The integrand is well behaved when the denominator
$\omega-\vn\cdot\vJ'$ vanishes.  That means that we may excise the
plane $\omega-\vn\cdot\vJ'=0$ from the integration range without changing
the value of the integral.  So, carrying out the inverse Fourier
transform after splitting the integral into two produces
\begin{equation}
  \begin{split}
&    \frac 1n f_\vn(\vJ|\omega)
    =
    -\tilde\Phi_\vn\ell^2\frac1{\omega-\vn\cdot\vJ}\vn\cdot\frac{\partial F}{\partial\vJ}
\\
&
\quad+\left[  f^\perp_\vn(\vJ_\perp|\omega)
      +\tilde\Phi_\vn(\omega)\ell^2
      \PV\int\frac{\d^3\vJ'}{\omega-\vn\cdot\vJ'}
      \right]
      \delta\left({\vn\cdot\vJ-\omega}\right),
\\
&
  \end{split}
\end{equation}
where $\PV$ denotes the Cauchy principal value.  Up to a
multiplicative factor this agrees with the expression for the DF of a
van Kampen mode given by equation (4) of Box~5.1 of
\cite{BinneyGalacticDynamicsSecond2008}.

The van Kampen modes~\eqref{eq:vanKFT} do not satisfy the dispersion
relation~\eqref{eq:dispersionreln}.  The reason for this is simple,
although possibly not immediately obvious: we obtained the dispersion
relation from the integral equation~\eqref{eq:Phiresp}, which in turn
came from integrating the (Fourier-transformed) CBE~\eqref{eq:kCBE}
with an implicit assumption that $\tilde f_\vn(\vk)\to0$ as
$|\vk|\to\infty$.  The van Kampen modes~\eqref{eq:vanKFT} do not
vanish as $|\vk|\to\infty$ and therefore do not satisfy the dispersion
relation.  For the same reason they are singular when viewed in action
or velocity space.  The relationship between waves that satisfy the
dispersion relation (the Landau modes) and van Kampen modes is
discussed in \cite{LauStellarfluctuationsstellar2021}.

\begin{figure*}
  \includegraphics[width=0.4\hsize]{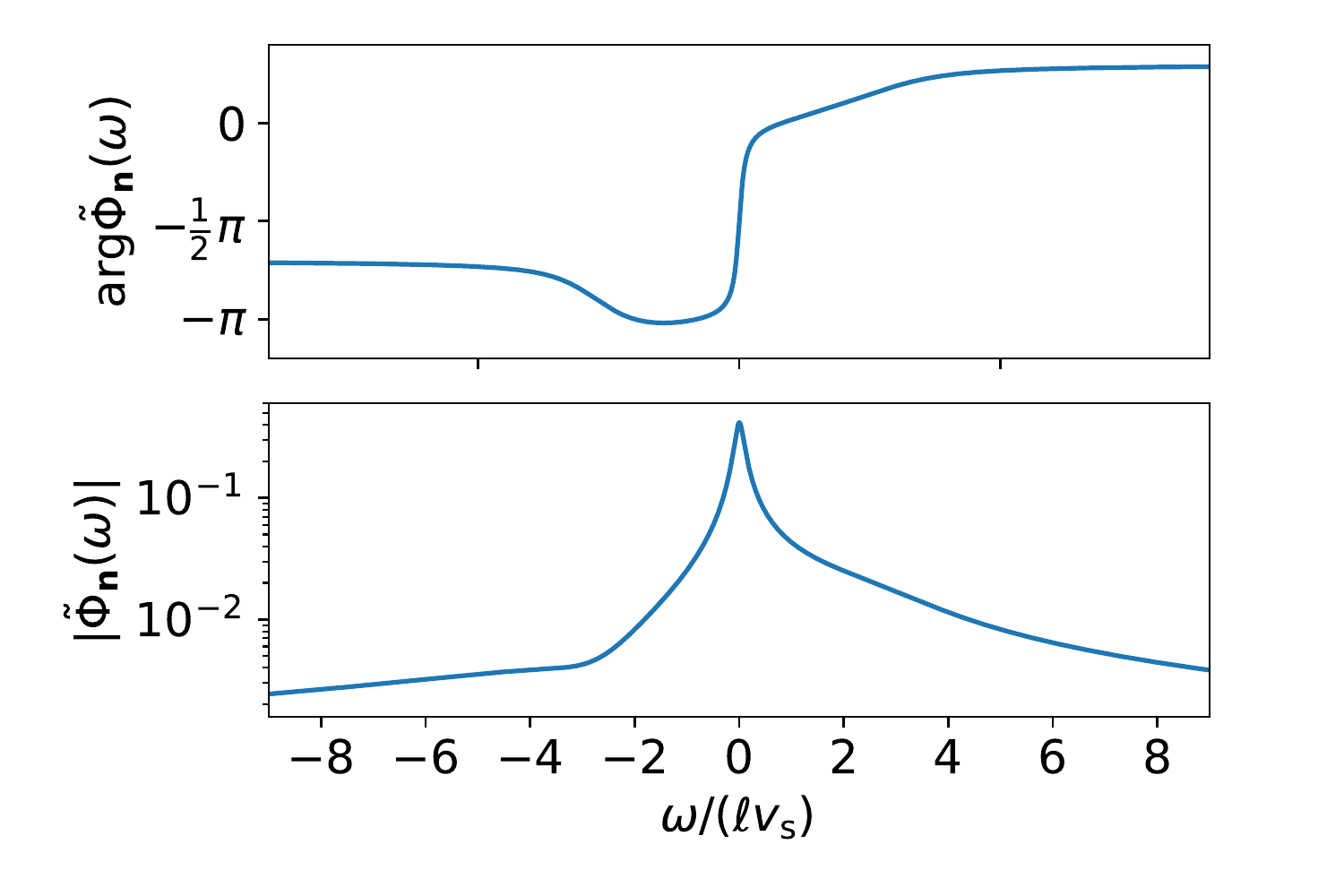}
  \includegraphics[width=0.4\hsize]{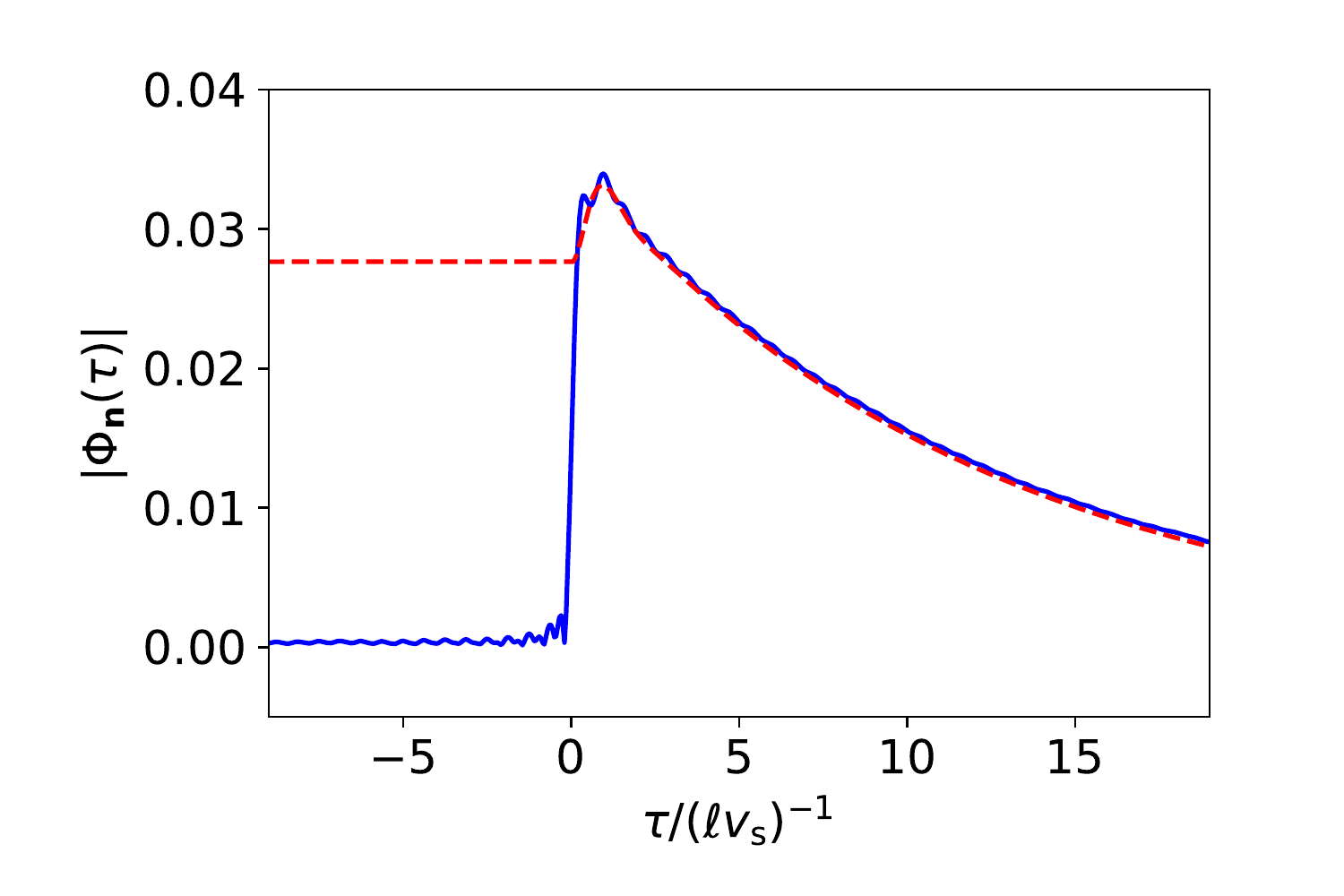}
  \caption{{\bf Left:} Spectrum $\tilde\Phi_\vn^+(\omega)$
    (equation~\ref{eq:vanKwake}) of the van Kampen-mode expansion of
    the $\vn=(0,0,1)$ component of the wake caused by a satellite
    moving with speed  $\vv\pert=(0,0,2)\vsc$ in a cube of mass $M=0.9M_{\rm J}$
    with a Maxwellian DF. {\bf Right:}
    the time-dependent potential $\Phi_\vn^+(\tau)$ reconstructed from modes
    having $|\omega|<10$ 
    (solid curve) compared to the numerical solution of the
    Volterra equation~\eqref{eq:Phiresp} (dashed curve) in which a
    slowly grown satellite is suddenly removed at $t=0$.
  }
  \label{fig:vanK}
\end{figure*}
Standard properties of Fourier transforms guarantee that any
well-behaved perturbation $f_\vn(\vk,\tau)$ or
$\tilde f_\vn(\vk,\tau)$ can be expressed as a superposition of van
Kampen modes~\eqref{eq:vanKFT}.  Here is how to perform this
decomposition given a snapshot $f_\vn(\vJ,\tau=0)$ at $\tau=0$.
Integrate the characteristic equation~\eqref{eq:charode} from this
$\tau=0$ initial condition to obtain
\begin{equation}
  \begin{split}
    \Phi_\vn(\tau)
    &
    =\frac{MV_\vn}{\ell^3}\tilde f_\vn(\vn \tau,0)\\
    &
    -\frac{MV_\vn}{\ell}
    \int_0^\tau\d\tau'\,\vn^2(\tau-\tau')\tilde F(\vn(\tau-\tau'))\Phi_\vn(\tau').
  \end{split}
\label{eq:Volterrat0}
\end{equation}
and split $\Phi_\vn(\tau)=\Phi^+_\vn(\tau)+\Phi^-_\vn(\tau)$, where $\Phi^+_\vn(\tau)$
is zero for $\tau<0$ and $\Phi^-_\vn(\tau)$ is zero for $\tau>0$.
Then~\eqref{eq:Volterrat0} becomes a pair of integral equations, one for
$\Phi^+_\vn(\tau)$ that holds only for $\tau\ge0$, the other for
$\Phi^-_\vn(\tau)$ that holds only for $\tau\le0$.  Introduce
the Fourier transforms
\begin{equation}
  \tilde\Phi_\vn(\omega)
  =
  \int_{-\infty}^\infty \e^{\i\omega\tau}\Phi_\vn(\tau)\,\d\tau
  ;
  \quad
  \Phi_\vn(\tau)=\frac1{2\pi}\int_{-\infty}^\infty
  \e^{-\i\omega\tau} \tilde\Phi_\vn(\omega)\,\d\omega.
\end{equation}
Fourier transforming the integral equation~\eqref{eq:Volterrat0}
for~$\Phi^+_\vn(\tau)$ gives
\begin{equation}
  \begin{split}
    \tilde\Phi^+_\vn(\omega)
    &=
    \frac{MV_\vn}{\ell^3}
    \int_{0}^\infty \e^{\i\omega\tau}\tilde f_\vn(\vn\tau,0)\,\d\tau
    +\pol_\vn(\omega)\tilde\Phi^+_\vn(\omega),
  \end{split}
  \label{eq:PhivanKa}
\end{equation}
where we have used $\tilde\Phi^+_\vn(\tau)=0$ for $\tau<0$ to extend
the lower limit of the integral in the second term
of~\eqref{eq:Volterrat0} from $\tau=0$ to $\tau\to-\infty$.
Rearranging, we have that $\Phi^+_\vn(\tau)$ is superposition of
modes $\tilde\Phi^+_\vn(\omega)\e^{-\i\omega\tau}$ with
amplitudes
\begin{equation}
  \begin{split}
    \tilde\Phi_\vn^+(\omega)
    &=\frac{2\pi}{1-\pol_{\vn}(\omega)}\frac{MV_\vn}{\ell^3}
    \int f_\vn(\vJ,0)\delta(\omega-\vn\cdot\vJ)\,\d^3\vJ.
  \end{split}
  \label{eq:PhivanK}
\end{equation}
The $\e^{-\i\omega\tau}$ time dependence of the modes means that
$\tilde\Phi_\vn^-(\omega)=[\tilde\Phi_\vn^+(\omega)]^\star$ and
therefore that
$\tilde\Phi_\vn(\omega)=\tilde\Phi^+_\vn(\omega)+\tilde\Phi_\vn^-(\omega)
=2\Re\tilde\Phi_\vn^+(\omega)$.

Simiarly, we can find $\tilde f_\vn(\vk_\perp|\omega)$ by first expressing our initial
condition as a superposition of modes,
\begin{equation}
  \tilde f_\vn(\vk,0)=\frac1{2\pi}\int\d\omega'\, \tilde f_\vn(\vk_\perp,k_\parallel|\omega').
\end{equation}
Multiplying this
by $\e^{\i\omega k_\parallel/n}$, integrating over $k_\parallel$ and
rearranging the result we have that
\begin{equation}
  \begin{split}
    \tilde f_\vn^\perp(\vk_\perp|\omega)
    &
    = \frac1n\int_{-\infty}^\infty \tilde
  f_\vn(\vk_\perp,k_\parallel,0)\e^{\i\omega k_\parallel/n}\d
  k_\parallel
  \\
  &\quad+\frac{\ell^3}{MV_\vn}\tilde\Phi_\vn(\omega)\pol_\vn(\omega),
  \end{split}
\end{equation}
which reduces to~\eqref{eq:PhivanKa} when $\vk_\perp=\vzero$.

For illustration, consider the steady-state wake produced by a
satellite of mass $M\pert$ that moves within the cube at constant
speed~$\vv\pert$ for $\tau<0$.  Suppose that the satellite has
arrived at $\vtheta=\vzero$ at time $\tau=0$, at which point it is
suddenly removed.  At that instant the perturbed DF of the cube
is given by equation~\eqref{eq:scresponsef} with $\Phi\stim_\vn=M\pert
V_\vn/(2\pi)^3$.  Substituting this $\tilde f_\vn(\vk,0)$
into~\eqref{eq:PhivanK} gives
\begin{equation}
\begin{split}
\tilde\Phi_\vn^+(\omega)
  &=
  -\frac{\i n^2M\pert V_\vn}{(2\pi)^3\ell^3}\frac1{1-\pol_\vn(\omega)}
  \frac1{1-\pol_\vn(\vn\cdot\vJ\pert)}\\
  &\quad\times
  \frac{\pol_\vn(\omega)-\pol_\vn(\vn\cdot\vJ\pert)}{\omega-\vn\cdot\vJ\pert}.
   \end{split}
\label{eq:vanKwake}
 \end{equation}
The left panel of Figure~\ref{fig:vanK} shows this
$\tilde\Phi_\vn^+(\omega)$ for $\vn=(0,0,1)$ in the case of a
satellite moving with $\vv\pert=(0,0,2)\vsc$ in a cube of mass
$M=0.9M_{\rm J}$ with a Maxwellian velocity distribution.  The
right panel plots the positive-time potential $\Phi^+_\vn(\tau)$ reconstructed
from~\eqref{eq:vanKwake} and compares to the results obtained from the
Volterra equation~\eqref{eq:Phiresp} in which the satellite grows
slowly over $-80<\tau/(\ell\vsc)^{-1}<-20$, has constant mass for $-20<\tau/(\ell\vsc)^{-1}<0$ and then is
removed at $\tau=0$.

\section{Quasi-linear evolution of an isolated $N$-body system}

\label{sec:LB}
So far we have treated the cube as a purely collisionless system.  Now
let us consider a cube composed of $N$ equal-mass stars, each of mass
$M/N$.  Each star will induce a wake as it passes through the cube.
Bearing in mind the caveats found in the time-dependent calculations
of Sections \ref{sec:volterrat} and~\ref{sec:damping} we can use the
results from Section~\ref{sec:steadystatewake} to calculate the net
response of the cube to all of these wakes.
We model the evolution of the $N$-body system as a superposition of
uncorrelated, dressed two-particle interactions, an idea that goes
back to
\cite{RostokerSuperpositionDressedTest1964,RostokerTestParticleMethod1964}:
see \cite{Hamiltonsimpleheuristicderivation2021} for a lucid overview
of the history of this idea, together with another example of its
application to the more general case of inhomogeneous systems \citep{HeyvaertsBalescuLenardtypekinetic2010,ChavanisKinetictheorylongrange2012}.

We assume that the cube is isolated and not subject to any externally
imposed perturbation.  Instead, the fluctuations caused by shot noise
in the DF will act as an effective ``external'', stirring potential,
$\Phi\stim$.  That is, the discrete nature of the $N$-body realisation
gives rise to a perturbation
\begin{equation}
  \rho\stim(\vtheta,t)=\frac1N\sum_{n=1}^N\delta(\vtheta-\vtheta_n(t))-\rho_0,
  \label{eq:rhodelta}
\end{equation}
in the density of our idealized smooth, collisionless cube: in the
$N$-body system the smooth background density
$\rho_0=\frac M{(2\pi)^3}$ is replaced by a sum of point masses
located at the stars' positions.
We model the stars' motion as the straight-line orbits they would
follow in the unperturbed system.
Then the angles of our ensemble of $N$ stars vary with time as
$\vtheta_n(\tau)=\vJ_n\tau+\vtheta_{\vn,0}$.
The actions $\vJ_\vn$ are constant.
From equations \eqref{eq:Vnprop} and~\eqref{eq:rhodelta} or
from~\eqref{eq:stirpot} the stirring potential corresponding
to~\eqref{eq:rhodelta} is 
\begin{equation}
  \Phi\stim_{\vm}(\tau)=\frac{ MV_\vm}{(2\pi)^3\ell^3N}\sum_{n=1}^N
  \e^{-\i\vm\cdot\vtheta_n(\tau)}.
\end{equation}
We assume that the cube settles down quickly enough that we can
consider each star to be followed by a steady-state wake, as described
in Section~\ref{sec:steadystatewake}.  Using equation~\eqref{eq:scresponsePhi},
the full potential, including the contribution from the wakes, is
given by
\begin{equation}
  \Phi_{\vm}(\tau)=\frac{ M V_\vm}{(2\pi)^3\ell^3N}\sum_{n=1}^N
  \frac1
  {1-\pol_\vm(\vm\cdot\vJ_{n})}
  \e^{-\i\vm\cdot\vtheta_{n}(\tau)}.
  \label{eq:LBPhi1}
\end{equation}
Similarly, using \eqref{eq:DF1undressed} and \eqref{eq:scresponsef}, the perturbed DF is given by
\begin{equation}
  \begin{split}
    &\tilde f_{\vm}(\vk,\tau)=\frac1{(2\pi)^3N}\times\\
    &\sum_{n=1}^N\Bigg[
    \e^{-\i\vk\cdot\vJ_{n}}+
    \frac{\tilde\pol_\vm(\vm\cdot\vJ_{n},\vk)}{1-\pol_\vm(\vm\cdot\vJ_{n})}\Bigg]
    \e^{-\i\vm\cdot\vtheta_{n}(\tau)}.\\
  \end{split}
  \label{eq:LBDF1}
\end{equation}
The first term in the square brackets comes from the contribution
$\delta(\vtheta-\vtheta_n)\delta(\vJ-\vJ_n)$ that each star makes
directly to $f(\vtheta,\vJ,\tau)$, the second from the wake that it
induces.  The perturbed DF $f(\vtheta,\vJ,\tau)$ also has a constant
$-F(\vJ)$ term, but that affects only the $\vm=(0,0,0)$ contribution to
the Fourier expansion of~$f$.

Armed with these expressions for the linearized response we can now
use the (nonlinear) characteristic evolution
equation~\eqref{eq:charode} to estimate the effect that this shot
noise has on the angle-averaged response $f_{(0,0,0)}(\vJ,\tau)$ over
many dynamical times.  We are more interested in the general
properties of this evolution than in the details for any particular
realisation of the stellar distribution.  So we
write~\eqref{eq:charode} as
\begin{equation}
  \label{eq:LBfundamentalk}
  \frac{\partial}{\partial t}\langle \tilde f_{(0,0,0)}(\vk,\tau)\rangle
  =
  -\sum_\vm\vm\cdot\vk
  \left\langle
    \tilde f_{-\vm}(\vk,\tau)\Phi_\vm(\tau)\right\rangle,
\end{equation}
where $\langle\cdot\rangle$ denotes an ensemble average, obtained by
averaging over all possible configuration histories of the system.
Taking the inverse Fourier transform of this expression yields a
continuity equation for $f_{(0,0,0)}(\vJ,t)$,
\begin{equation}
  \label{eq:LBfundamentalJ}
  \frac{\partial}{\partial t}\langle f_{(0,0,0)}(\vJ,t)\rangle
  =
  -\frac{\partial}{\partial\vJ}\cdot\vF(\vJ,t),
\end{equation}
in which the flux is given by
\begin{equation}
  \vF(\vJ,t)\equiv
  -\i  \sum_\vm\vm
  \left\langle
    f_{-\vm}(\vJ,t)\Phi_\vm(t)\right\rangle.
  \label{eq:LBfluxprimitive}
\end{equation}
Expressions \eqref{eq:LBfundamentalk}--\eqref{eq:LBfluxprimitive}
would be exact if we were to substitute the nonlinearly evolved
$f_\vm(\vJ,\tau)$ and~$\Phi_\vm(\tau)$ in the right-hand sides.  We
only have the linearized approximations~\eqref{eq:LBDF1}
and~\eqref{eq:LBPhi1} though, in which case the
flux~\eqref{eq:LBfluxprimitive} becomes a superposition of
$O(1/N^2)$ contributions from uncorrelated dressed two-body interactions.

To carry out the ensemble averaging in~\eqref{eq:LBfluxprimitive} we
use the following result from Monte Carlo integration.  Suppose that
points $x_1,...,x_N$ are drawn independently from a pdf $F(x)$ and let
$g(x)$ and $h(x)$ be functions of~$x$ whose means are zero.
Then the ensemble average
\begin{equation}
  \begin{split}
&  \left\langle
    \sum_{i=1}^N g(x_i)\sum_{j=1}^N h(x_j)
  \right\rangle
  \\
&=  \left\langle
    \sum_{i=1}^N g(x_i)h(x_i)
  \right\rangle
  +
    \left\langle
    \sum_{i=1}^N g(x_i)\sum_{j\ne i} h(x_j)
  \right\rangle
\\
&  = N\int \d x\, F(x)g(x)h(x),
  \end{split}
\end{equation}
the second term in the second line vanishing because the $x_i$ are independently drawn
from~$F$.
The ensemble averages on the RHS of~\eqref{eq:LBfundamentalk} then
become
\begin{equation}
  \begin{split}
    \langle\tilde f_{-\vm}\Phi_\vm\rangle
    &=
    \frac{MV_\vm}{(2\pi\ell)^3N}
    \int\d^3\vJ'
    F(\vJ')
    \frac{\e^{-\i\vk\cdot\vJ'}}{1-\pol_\vm(\vm\cdot\vJ')}
    \\
&
    +\frac{MV_\vm}{(2\pi\ell)^3N}
    \int\d^3\vJ'
    F(\vJ')
    \frac{\tilde\pol_{-\vm}(-\vm\cdot\vJ',\vk)}
    {|1-\pol_\vm(\vm\cdot\vJ')|^2},
  \end{split}
\end{equation}
where the denominator in the
second term comes from using the relation
$\pol_{-\vm}(-\omega)=\pol^\star_\vm(\omega)$ that follows directly
from the definition~\eqref{eq:In}.  Taking the inverse Fourier
transform of this expression and substituting the resulting
$\langle f_{-\vm}\Phi_\vm\rangle$ into~\eqref{eq:LBfluxprimitive}, the flux
can be split into two parts, $\vF=\vF_1+\vF_2$, in which 
\begin{equation}
  \label{eq:LBflux1}
  \vF_1=-\i\frac{ 1}{(2\pi\ell)^3N}F(\vJ)\sum_\vm\vm
  \frac{MV_\vm}{1-\pol_\vm(\vm\cdot\vJ)},
\end{equation}
is a ``drift'' term that depends linearly on $F(\vJ)$, and, using
equation~\eqref{eq:scresponsef},
\begin{equation}
  \label{eq:LBflux2}
  \begin{split}
    &  \vF_2=-\frac{\pi}{(2\pi\ell)^3N\ell}\times
    \\
    &\sum_\vm\vm
  |MV_\vm|^2
    \vm\cdot\frac{\partial F(\vJ)}{\partial \vJ}
    \int\frac{F(\vJ')\delta(\vm\cdot(\vJ-\vJ'))}
    {|1-\pol_\vm(\vm\cdot\vJ')|^2}\d^3\vJ'
  \end{split}
\end{equation}
is a ``diffusive'' term that depends on $\partial F/\partial\vJ$.

We may turn the expression for~$\vF_1$ into something that has the same
form as that for $\vF_2$ by rewriting the sum in~\eqref{eq:LBflux1} as
\begin{equation}
  \begin{split}
&    \sum_\vm\vm\frac{MV_\vm}{1-\pol_\vm(\vm\cdot\vJ)}
  =
  \sum_\vm
  \vm MV_\vm\frac{1-\pol_\vm^\star(\vm\cdot\vJ)}{|1-\pol_\vm(\vm\cdot\vJ|^2}\\
  &
  =\frac12\sum_\vm\vm MV_\vm
  -\frac{\pol_\vm^\star(\vm\cdot\vJ)-\pol^\star_{-\vm}(-\vm\cdot\vJ)}
  {|1-\pol_\vm(\vm\cdot\vJ)|^2}\\
  &
  =\frac12\sum_\vm\vm MV_\vm
  \frac{MV_\vm^\star\vm^2\left[\int_{-\infty}^\infty \tau \e^{\i\vm\cdot\vJ\tau}\tilde\fzero(\vm\tau)\d\tau\right]^\star}
  {\ell|1-\pol_\vm(\vm\cdot\vJ)|^2},\\
\end{split}
\label{eq:LBflux1tmp}
\end{equation}
where to obtain the last line we have used the definition~\eqref{eq:In} of $\pol_\vm(\omega)$.
Now notice that the integral in the square brackets can obtained by differentiating the equation
\begin{equation}
  \int_{-\infty}^\infty
  \e^{\i\vm\cdot\vJ\tau}\tilde\fzero(\vm\tau)\,\d\tau
  ={(2\pi)}\int \fzero(\vJ')\delta(\vm\cdot(\vJ-\vJ'))\d^3\vJ'.
\end{equation}
Using this in~\eqref{eq:LBflux1tmp} and substituting back
into~\eqref{eq:LBflux1} we obtain
\begin{equation}
  \vF_1=\frac{\pi}{(2\pi\ell)^3N\ell}\sum_\vm\vm|MV_\vm|^2
  F(\vJ)\int \frac{\vm\cdot\frac{\partial
      F}{\partial\vJ'}\delta(\vm\cdot(\vJ-\vJ'))}
  {|1-\pol_\vm(\vm\cdot\vJ')|^2}\d^3\vJ'.
\end{equation}
This can be combined with our expression~\eqref{eq:LBflux2} to write
the total flux as
\begin{equation}
  \begin{split}
&  \vF=
  \frac{\pi}{(2\pi\ell)^3N\ell}\sum_\vm\vm|MV_\vm|^2
\times\\
&
\int
  \frac{\delta(\vm\cdot(\vJ-\vJ'))}
  {|1-\pol_\vm(\vm\cdot\vJ')|^2}
\vm\cdot\left[
      F(\vJ)\frac{\partial F(\vJ')}{\partial\vJ'}
      -F(\vJ')\frac{\partial F(\vJ)}{\partial\vJ}\right]
  \d^3\vJ'.
  \end{split}
\label{eq:LBflux}
\end{equation}
Replacing $\vJ$ by $\ell\vv$ (equation~\ref{eq:aa}) and taking the
limit $\ell\to\infty$ of an infinite box, this action flux agrees with the
velocity flux for an infinite homogeneous system given by equation~(36)
of \cite{ChavanisKinetictheoryspatially2012}, once we account for the
different normalisations used for the DF $F$ and the potential $V_\vm$.

\begin{figure}
  \includegraphics[width=\hsize]{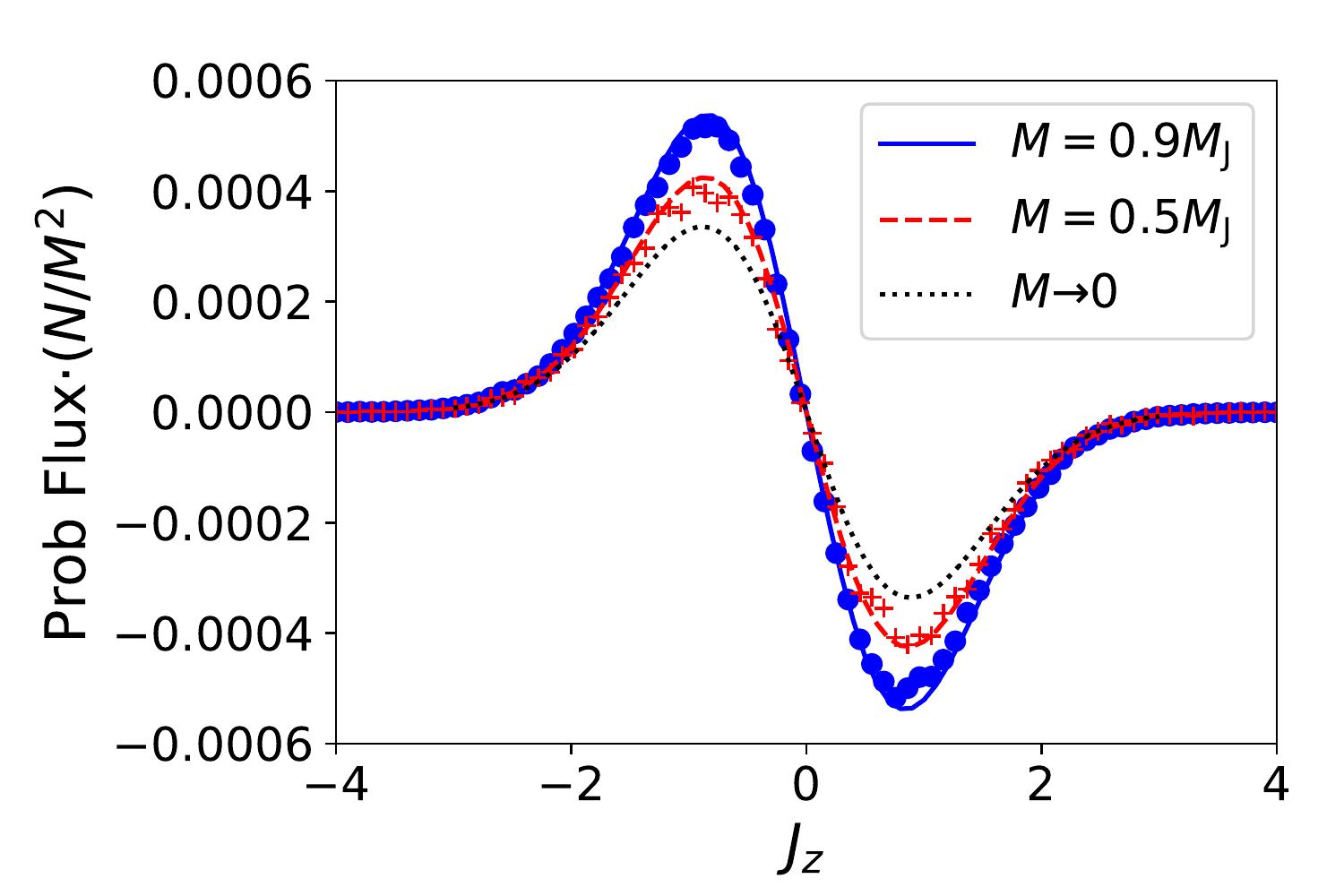}
  \caption{Comparison of the Lenard-Balescu flux
    predictions~\eqref{eq:LBfluxexample} against those
    measured from ensembles of $N$-body simulations for cubes with
    DF~\eqref{eq:LBtestDF} and masses $M=0.9M_{\rm J}$ and $0.5M_{\rm
      J}$.  The fluxes are scaled by $N/M^2$ to isolate the effects of
    the enhancement from particle dressing.  The $M\to0$ curve
    (dotted) shows the flux obtained when this dressing is turned off. }
  \label{fig:LB}
\end{figure}

\subsection{Test}

The flux~\eqref{eq:LBflux} vanishes if $F(\vJ)$ is Maxwellian.  A more
interesting result occurs if we take $F(\vJ)$ to be a sum of
Maxwellians though.  Writing
\begin{equation}
 G(\vJ|\vJ_0)=\frac1{(2\pi)^{3/2}}\exp\left[-\frac12(\vJ-\vJ_0)^2\right]
\end{equation}
for a unit-dispersion Maxwellian centred on $\vJ=\vJ_0$, we take as
our DF
\begin{equation}
 F(\vJ)=\frac1{2(2\pi)^3}\left[G(\vJ|\vJ_0)+G(\vJ|-\vJ_0)\right].,
 \label{eq:LBtestDF}
\end{equation}
a superposition of two Maxwellians, one centred on
$\vJ_0=(0,0,\frac12)$, the other at $-\vJ_0$ (see also \cite{ChavanisDynamicalstabilityinfinite2012}).
The Lenard--Balescu flux~\eqref{eq:LBflux} for this DF is
\begin{equation}
  \begin{split}
  \vF=&
\frac{\pi}{(2\pi\ell)^3N\ell}
\sum_\vm\frac{\vm|MV_\vm|^2}{|1-\pol_\vm(\vm\cdot\vJ)|^2}
\times\\
&
\frac{\vm\cdot\vJ_0}{2(2\pi)^6}\left[
  G(\vJ|-\vJ_0)G_\vm(\vJ|\vJ_0)
  -
  G(\vJ|\vJ_0)G_\vm(\vJ|-\vJ_0)
\right],
\end{split}
\label{eq:LBfluxexample}
\end{equation}
in which
\begin{equation}
  \begin{split}
G_\vm(\vJ|\vJ_0)&\equiv  \int\d^3\vJ'\,\delta(\vm\cdot(\vJ-\vJ'))
  G(\vJ|\vJ_0)
\\
  &  =
  \frac1{\sqrt{2\pi\vm^2}}
  \exp\left[-\frac{(\vm\cdot(\vJ-\vJ_0))^2}{2\vm^2}\right].
  \end{split}
\end{equation}
The curves on Figure~\ref{fig:LB} plot the $z$-component of this flux
for cubes of mass $0.5M_{\rm J}$ and $0.9M_{\rm J}$ in which the
contribution to the potential of spatial modes beyond $n_{\rm max}=4$
is suppressed.  The fluxes are scaled by $N/M^2$ in this plot so that
the differences between them are due to the dressing effects of
self-gravity (the $|1-\pol_\vm|^2$ denominator in~\eqref{eq:LBflux}).
For reference the $M\to0$ case, in which self-gravity is switched off
($\pol_\vm=0$), is plotted as as the dotted curve.  The $|\vm|=1$
spatial modes are by far the dominant contributors to this enhancement
by self gravity.

Measuring Lenard--Balescu fluxes from $N$-body simulations is
expensive \citep{LauRelaxationsphericalstellar2019}: one has to run
each simulation for at least a few dynamical times to allow
the particle dressing to become established, but not so long that the
DF relaxes significantly.  We address this by running 2500 simulations
each with $N=10^5$ particles for $\tau=500/\ell\vsc$ time units.  This
is of order
100 dynamical times, but the large $N$ means that it is only a tiny
fraction of a relaxation time: the mean axis ratio of the velocity
ellipsoids decays from its initial value of about 1.250 to 1.248
($M=0.5M_{\rm J}$) or 1.242 ($M=0.9M_{\rm J}$) over the course of the
runs.  To measure the fluxes we set up a regularly spaced stack of
constant-$J_z$ planes and simply accumulate the net flux through each
plane at every $\Delta\tau=0.01/\ell\vsc$ timestep.

The points in Figure~\ref{fig:LB} plot the fluxes measured from the
simulations.  The agreement with the analytical predictions is good,
although the simulated fluxes tend to be slightly smaller in
magnitude; this is most probably caused by the small amounts of
relaxation in the simulations that are not accounted for in the
analytical calculation.

\section{Conclusions}

We have used linear perturbation theory to calculate the response of
stars within a periodic cube to a satellite passing through them.
This is an extremely idealised problem, which raises the
question of to what extent our findings are applicable to more
realistic galaxy models.
In order of increasing usefulness, we note that:
\begin{enumerate}
\item The most unusual feature of our approach is the simplification
  afforded by Fourier transforming the DF in~$\vJ$.  This is possible
  only because the orbital frequency vector $\vOmega(\vJ)$ in the
  periodic cube is a linear function of $\vJ$.  In more realistic
  galaxy models the relation between $\vOmega$ and $\vJ$ is not linear
  (at least not globally), which means that the Fourier-transformed
  CBE is no longer a first-order PDE and therefore cannot be solved
  using the method of characteristics.
\item Poisson's equation for the cube separates in angle-action
  variables.  More general systems do not usually admit such a
  convenient separation, but, on the other hand, there is a well-known
  remedy: the matrix method of \cite{KalnajsDynamicsFlatGalaxies1976}.
\item One can use Kalnajs' matrix method to write down a generalized
  version of the Volterra integral equation~\eqref{eq:Phiresp} for the
  evolution of the potential
  \citep[e.g.,][]{MuraliTransmissionAmplificationDisturbances1999,Dootson}
  as an explicit function of time, the natural starting point for
  solving realistic initial-value problems.  Having
  obtained this equation for the cube the dispersion relation drops
  out immediately (Section~\ref{sec:linstab}).  More interestingly,
  when solving the Volterra equation it is straightforward to include
  the backreaction of the response on any external perturbation (e.g.,
  Section~\ref{sec:damping}).
\item Stellar systems are most fascinating when they are most
  responsive
  \citep[e.g.,][]{JulianNonAxisymmetricResponsesDifferentially1966,Binneyshearingsheetswing2020}.
  The cube becomes interesting when its mass approaches the Jeans mass
  and the amplification factors $(1-\pol_\vn)^{-1}$ of the
  longest-wavelength $|\vn|=1$ modes become large.
\item It is straightforward to decompose any perturbation of the
  cube into its underlying van Kampen modes, offering the
  possibility of using it as a test bed to understand thermal
  fluctuations in stellar systems
  \citep{LauStellarfluctuationsstellar2021}.
\item
  Unsurprisingly, we find that predictions from linear theory agree
  very well with measurements from $N$-body simulations when the
  perturbations are small, such as in the calculations of the
  Lenard--Balescu flux for $N=10^5$-body systems
  (Figure~\ref{fig:LB}).
  What is perhaps more surprising is how quickly linear theory becomes
  suspect as the perturbation amplitude increases.  For example, our
  calculation of the dynamical friction on a massive satellite
  (Figures \ref{fig:volterradrag} and~\ref{fig:damp}) is only a
  qualitative match to the $N$-body results for satellite masses as
  small as $10^{-2}\,M$.
  The key ingredient missing from linear theory is the
  ``capture'' of stars by the satellite: in the perturbed cube the
  orbits of a significant number of stars become bound to the
  satellite; they are not the slightly perturbed straight-line
  orbits as assumed by linear theory.
  \end{enumerate}

\section*{Acknowledgments}

I thank James Binney, Rimpei Chiba, Jean-Baptiste Fouvry, Jun Lau and
Christophe Pichon for insightful comments and encouragement.  This
work was supported by the UK Science and Technology Facilities Council
under grant number ST/S000488/I.

\section*{Data availability}
No new data were generated or analysed in support of this research.

\bibliographystyle{mnras}

\appendix

\section{$N$-body simulations}
\label{sec:Nbody}

The periodic cube is straightforward to model by $N$-body simulation.
We use $N$ particles to sample the cube's initial, equilibrium DF
$\fzero(\vx,\vv)$, assigning each a mass of $M/N$.  When a perturbing
satellite is added, it is treated as an additional particle of
mass~$M\pert$.

There is a single length scale and a single time scale.  The regular,
periodic nature of the cube means that it is natural to model the
density field $\rho(\vx)$ and potential $\Phi(\vx)$ using a regular
cubic mesh.  Then $\Phi(\vx)$ can be obtained from $\rho(\vx)$ using a
fast Fourier transform on the mesh of $\rho(\vx)$ values to obtain
$\rho_\vn$, multiplying the result by $V_\vn$ (equation~\ref{eq:Vn})
to obtain $\Phi_\vn$ and then fast Fourier transforming back to obtain
the values of $\Phi(\vx)$ on the mesh.
We use a nearest grid-point scheme to assign particles' masses to the
$\rho(\vx)$ mesh and the same scheme, coupled with second-order finite
differencing, to assign accelerations to the particles from the
$\Phi(\vx)$ mesh.  The particles' phase-space coordinates are evolved
using a drift--kick--drift leapfrog integrator with timestep
$\Delta\tau=0.01/\ell\vsc$.

For most of the calculations presented here we include spatial modes
up to $n_{\rm max}=18$ in the potential: modes of higher order than
this will be produced in the density field from nonlinear interactions
among lower-order modes, but they will have no effect on the
potential.  We use meshes of $(8n_{\rm max})^3=144^3$ points to
represent the density and potential, the factor of 8 a compromise
between having true translation invariance (which would require an
infinite number of mesh points) and computational feasibility.

\end{document}